\documentclass[%
 aip,
 amsmath,amssymb,
 reprint,%
]{revtex4-1}

\usepackage{graphicx}
\usepackage{dcolumn}
\usepackage{bm}
\usepackage{color}

\usepackage[utf8]{inputenc}
\usepackage[T1]{fontenc}
\usepackage{mathptmx}
\usepackage{etoolbox}

\makeatletter
\def\@email#1#2{%
 \endgroup
 \patchcmd{\titleblock@produce}
  {\frontmatter@RRAPformat}
  {\frontmatter@RRAPformat{\produce@RRAP{*#1\href{mailto:#2}{#2}}}\frontmatter@RRAPformat}
  {}{}
}%
\makeatother
\begin{document}
 
\preprint{AIP/123-QED}

\title{Chaos aided regime of Laser/Electromagnetic Energy Absorption by plasma}
\author{Rohit Juneja$^*$}
\email{onlyforjuneja@gmail.com}
\author{Amita Das$^*$}
 \email{amita@iitd.ac.in}
 \author{Trishul Dhalia}%
 \author{Animesh Sharma}
 \affiliation{ 
Department of Physics, Indian Institute of Technology Delhi, Hauz Khas, New Delhi 110016, India}

\date{\today}

\begin{abstract}
\textbf{Abstract}\\
The absorption of laser energy by plasma is of paramount importance for various applications.  Collisional and resonant processes are often invoked for this purpose. However, in some contexts (e.g. in vacuum \cite{brunel1987not} and the $\vec{J} \times  \vec{B}$ heating \cite{kruer1985j}), the energy transfer occurs even when plasma is collisionless, and there is no resonant process involved. The energy absorption in these cases has been attributed to the sheath electrostatic fields that get generated as the electrons are pulled out in the vacuum from the plasma medium. The origin of irreversibility aiding the absorption, in these cases, remains to be understood.  Particle-In-Cell (PIC) simulations using the OSIRIS 4.0 platform have been carried out. The nearby trajectories of lighter electron species involved in the interaction with the laser show exponential separation.  This is confirmed by the positive Lyapunov index and also by other characterizations. The observations in these cases are contrasted with the electron cyclotron resonant case, which shows negligible chaos in the electron trajectories despite the energy absorption percentage being high.   \\


\end{abstract}

\maketitle

\begin{quotation}
A detailed study of electron trajectories generated from the Particle-In-Cell (PIC) simulations for laser-plasma interaction has been carried out. It is shown that when the laser energy absorption gets mediated by the generation of electrostatic fields, the electron trajectories are chaotic as depicted by the positive value of the Lyapunov index. On the other hand, when the process is governed by the resonant mechanism and is in no way mediated by the generation of electrostatic fields, the electron trajectories are regular.  The origin of irreversible energy transfer in the former case can thus be traced to the chaotic nature of the electron trajectories. 

\end{quotation}

\section{\label{intro}Introduction} 
The transfer of laser energy to the plasma medium is crucial in the context of many applications. These include the fast ignition scheme of laser fusion, laser-based particle acceleration schemes, etc.,\cite{kaw1969laser, kaw2017nonlinear, das2020laser}. This has led to research studies towards understanding the mechanisms underlying the process of irreversible energy transfer from laser/ electromagnetic (EM) waves to the plasma medium. There are many well-known collisional and collisionless mechanisms for the coupling of laser energy with plasma medium which have been studied and discussed in detail in literature \cite{kruer2019physics,decker1994nonlinear,estabrook1975two,freidberg1972resonant, ping2008absorption, wilks1992absorption, chopineau2019identification, stix1965radiation, gibbon1992collisionless}.
A single charged particle can irreversibly gain energy from an oscillatory electromagnetic field provided the  $\langle \vec{\mathrm{v}} \cdot {\vec{E}} \rangle$ is finite. Here the angular brackets indicate the time-averaging process.  This requires that the velocity and the electric field should be oscillating in phase. From the  force  equation below ( Eq.(\ref{eq:Force-E}))
\begin{equation}
    \frac{d\vec{\mathrm{v}} }{dt} = \frac{q}{m} \vec{E}(\vec{r},t) 
    \label{eq:Force-E}
\end{equation}
the velocity of the particle $\vec{\mathrm{v}}$ will always be out of phase with the electric field $\vec{E}$ and hence $\langle \vec{\mathrm{v}} \cdot {\vec{E}} \rangle = 0$.  This rules out any net energy transfer from the oscillating electric field to any charged particle.  However, if the particle also undergoes collision,   the required phase shift for net energy gain by the particle may be introduced. 

For a collection of charged particles in plasma,  instead of the particle velocity, the condition applies on the current in the medium, i.e., $\langle \vec{J} \cdot \vec{E} \rangle$ needs to be finite, where $\vec{J} = e(n_i\vec{\mathrm{v}}_i - n_e \vec{\mathrm{v}}_e) $.  Here $e$ is the electron charge and $n_s, \mathrm{v}_s$ are the density and velocity respectively, of the species $s = e,i$ (representing electrons and ions).  Again the presence of collisions/ resistivity in the medium can create an in-phase component of current with the applied electric field for a finite and irreversible energy transfer. The energy transfer processes relying on collisions are known as collisional/resistive processes.  The plasma medium being a collection of particles, also supports a variety of collective modes in the form of oscillations and waves. When the EM/ laser frequency matches the frequency of these waves, resonance absorption occurs. For these cases, collisions/resistivity can become inconsequential. There exists, however, additional absorption mechanisms that neither rely on collisional nor on resonant processes  \cite{brunel1987not,kruer1985j}. In these cases, it is observed that if a component of the electromagnetic force is normal to the plasma vacuum boundary, it can pull the electrons from the plasma in the vacuum region (the lighter electron species of the plasma are the ones that dominantly participate in the dynamics at the fast laser time scales). The combination of the sheath plasma field and the electromagnetic fields is believed to aid the absorption process. However, the underlying mechanism for irreversible energy transfer is not clearly understood in these cases. We demonstrate using Particle-In-Cell (PIC) simulations that in such situations, the nearby electron trajectories show exponential separation, and the Lyapunov index is found to be positive. The role of chaotic electron trajectories is ascertained for the non-resonant collisionless cases by demonstrating that for higher energy absorption, the value of the Lyapunov index is higher. It is also shown that the electron trajectories remain regular when the resonant process is operative.  

It should be mentioned here that the phenomena of chaos have also been observed in many different contexts of studies carried out for the plasma medium \cite{valvo2022numerical,samanta2021comparative,torres2023study,deshwal2022chaotic,nurujjaman2007parametric,roy2023chaos,sonnino2020nonlinear}. Some of these studies are related to understanding the diffusive transport of charged particles in the presence of magnetic fields \cite{valvo2022numerical,samanta2021comparative}. In the studies conducted by Samanta et. al.,  \cite{samanta2021comparative}, the magnetic field was chosen as spatially chaotic. The question of turbulent transport in the context of magnetized plasma for Tokamaks was modeled and analyzed with the help of Maps by Torres {\it{et al.}} \cite{torres2023study}. The presence of chaotic dynamics for small-sized dust clusters obeying Yukawa interaction was demonstrated recently \cite{deshwal2022chaotic}. There have also been experimental observations of chaos. For instance, in the context of DC glow discharge, certain self-excited oscillations associated with anode glow showed chaotic dynamics \cite{nurujjaman2007parametric}. 

We have employed here a comprehensive Particle-In-Cell (PIC) simulation study for a laser/ EM wave interacting with a plasma medium. We demonstrate that the electron trajectories become chaotic cases where the energy absorption process is governed by the non-resonant collisionless processes. 
The paper has been organized as follows. In section \ref{sim}, we describe the simulation geometry and other details related to it.  In section \ref{JcrossB}, the simulation has been performed for the non-resonant cases, namely the  $\vec{J} \times \vec{B}$  and the vacuum heating mechanism. 
  
These observations have been contrasted with the studies conducted for the case where the electrons acquire energy from the laser field through the electron cyclotron resonance process in section  \ref{ECresonance}.  

In section \ref{diffusion}, it is shown that the energy acquired by the electrons ultimately gets thermalized through a diffusive process in the velocity space. 
Section \ref{conclusion} provides a conclusion of our findings.

\begin{table}
\caption{\label{tab:table1}Simulation parameters in normalized units and possible values in standard units.}
\begin{ruledtabular}
\begin{tabular}{lcc}
\textbf{Parameters}&\textbf{Normalised value}& \textbf{Value in standard units}\\
\hline\\
\hline
{\textbf{Laser Parameters}} \\ 
\hline\\
Frequency ($\omega_{L}$)  & $0.2 \omega_{pe} $ & $ 0.2 \times 10^{15} Hz$\\
Wavelength ($\lambda_{L}$)  & 31.4$c/\omega_{pe}$ &9.42 $\mu m$\\
Intensity ($I_0$) in 1D case & $a_0 = 5.0$ & $3.85\times10^{17}Wcm^{-2}$\\
Intensity ($I_0$) in 2D case & $a_0 = 0.5$ & $3.85\times10^{15}Wcm^{-2}$\\\\
\hline
\textbf{Plasma Parameters} \\
\hline\\
Number density ($n_{0}$) & 1 & $3.15 \times 10^{20} cm^{-3}$\\
Electron Plasma \\Frequency ($\omega_{pe}$) & 1 & $10^{15} Hz$\\
Skin depth ($c/\omega_{pe}$)   & $1$ & $0.3$ $\mu m$ \\\\

\hline
      
       {\textbf{Simulation Parameters}} \\{\textbf{in 1D case}} \\
      \hline\\
      $L_x$ & $3000$ & $900 \mu m$ \\
      $dx$ & $0.05$ & $15 nm$ \\ 
    $dt$ & $0.02$ & $ 2 \times 10^{-17} s$ \\ \\

    \hline
    {\textbf{Simulation Parameters}} \\{\textbf{in 2D case}} \\
      \hline\\
      $L_x$ & $1000$ & $300 \mu m$ \\
      $L_y$ & $1000$ & $300 \mu m$ \\
      $dx$ & $0.1$ & $30 nm$ \\
      $dy$ & $0.1$ & $30 nm$ \\
    $dt$ & $0.04$ & $ 4 \times 10^{-17} s$ \\ \\

    \hline 
      {\textbf{External Magnetic }} \\
      {\textbf{Field Parameters}} \\
      \hline\\
      $B_0$ (RL Mode) & $0.4$ & $2.27 kT$ \\
      $\omega_{ce}$ & $0.4$ & $0.4\times10^{15} Hz$ \\
      \hline

\end{tabular}
\end{ruledtabular}
\end{table}

\begin{figure*}
\includegraphics[scale = 0.26]{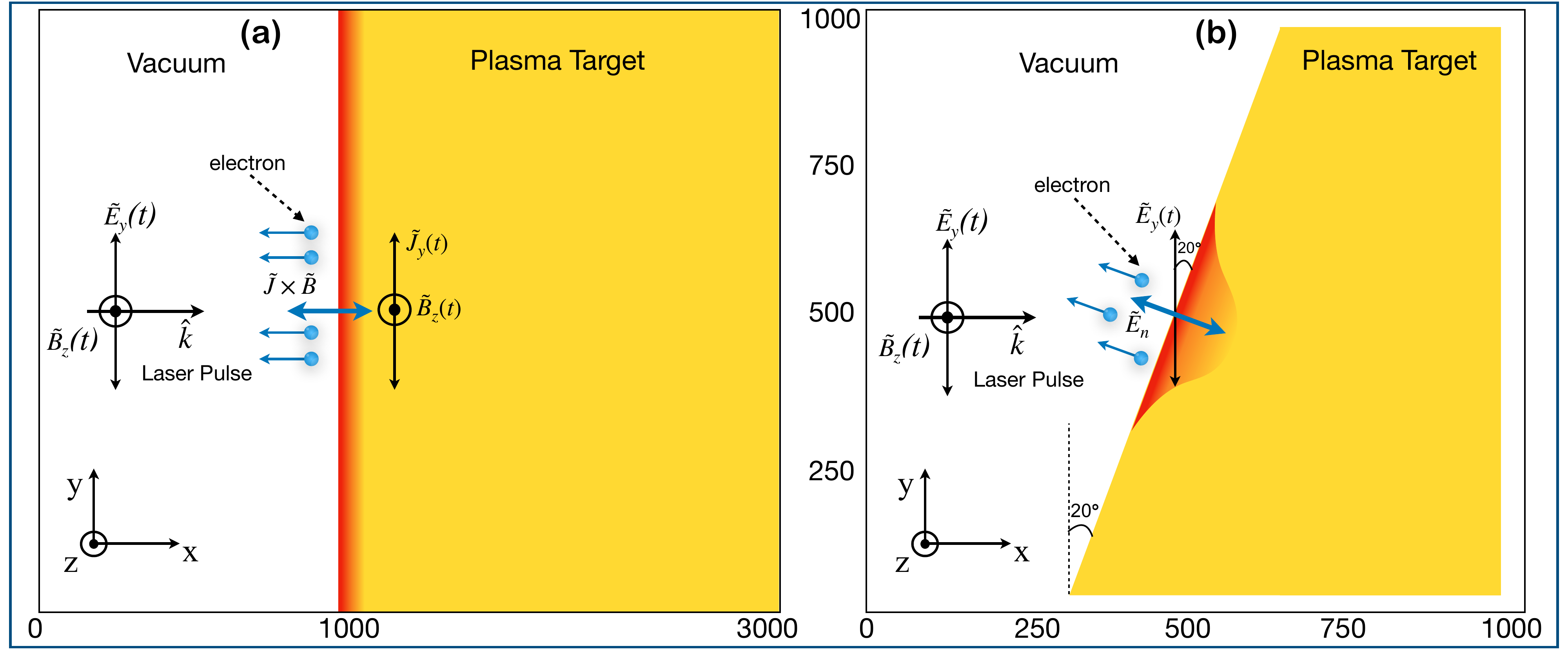}
\caption{\label{schematic} The schematic of laser pulse interacting with plasma for the configuration of  $\vec{J}\times\vec{B}$ heating has been shown in subplot (a).  The laser pulse, in this case, falls normal to the plasma target. When the laser intensity is relativistically high, the $\vec{J} \times \vec{B}$ force can extract electrons in and out of the target surface, creating sheath fields. In subplot (b), the configuration of vacuum heating has been shown schematically.  The laser is incident at an oblique angle and is p-polarised. The normal component of the laser electric field in this extracts the electrons in and out of the target. This generates sheath electric fields.}
\end{figure*}

\section{\label{sim}Simulation details}
We have carried out both one-dimensional (1D) and two-dimensional (2D) Particle-In-Cell (PIC) simulations using a massively parallel Particle-In-Cell code OSIRIS 4.0 \cite{hemker2000particle, fonseca2002osiris, fonseca2008one}. These studies are meant to understand the dynamical behavior of the electrons while different schemes of laser energy absorption are operative.  The schematic in Fig. \ref{schematic} shows the collisionless non-resonant schemes that we will study in section III.   Fig. \ref{schematic}(a), depicts a laser pulse that encounters an overdense plasma target at normal incidence. In this case, the direction of the electric field is normal to the surface and cannot extract the electrons in the vacuum. However, when the laser intensity is high such that the quiver velocity of the electrons within the skin depth of the laser becomes relativistic, the  $\vec{J} \times \vec{B}$ mechanism is operative. Here, $\vec{J}$ is the electron current due to the quiver motion of electrons, and $\vec{B}$ is the oscillating magnetic field of the laser. The $\vec{J} \times \vec{B}$ force is normal to the surface and is responsible for fetching the electrons in and out of the target, resulting in space charge development along the propagation direction of the laser. Thus, 1-D spatial variations suffice for studying this particular case as the spatial variations are only along the laser propagation direction.  However, the velocity and other fields have been chosen to have all three components. 

In Fig \ref{schematic}(b), the configuration of vacuum heating has been depicted. A p-polarised laser with non-relativistic intensity falls obliquely on the surface of the plasma target. The laser electric field is thus at an oblique angle, as depicted in the schematic figure. In this case, the normal component of the laser electric field to the plasma target surface can pull and push the electrons out and into the surface. 
This would result in the generation of electrostatic sheath fields, which, in conjunction with the electromagnetic field of the laser, are believed to be important for laser energy absorption. This mechanism is operative even at the non-relativistic intensity of the laser field. The oblique incidence, however, necessitates a 2-D simulation to observe this phenomenon. Several diagnostics to understand the behaviour of electron trajectories in these two cases have been performed. 

\begin{figure}
\includegraphics[scale = 0.26]{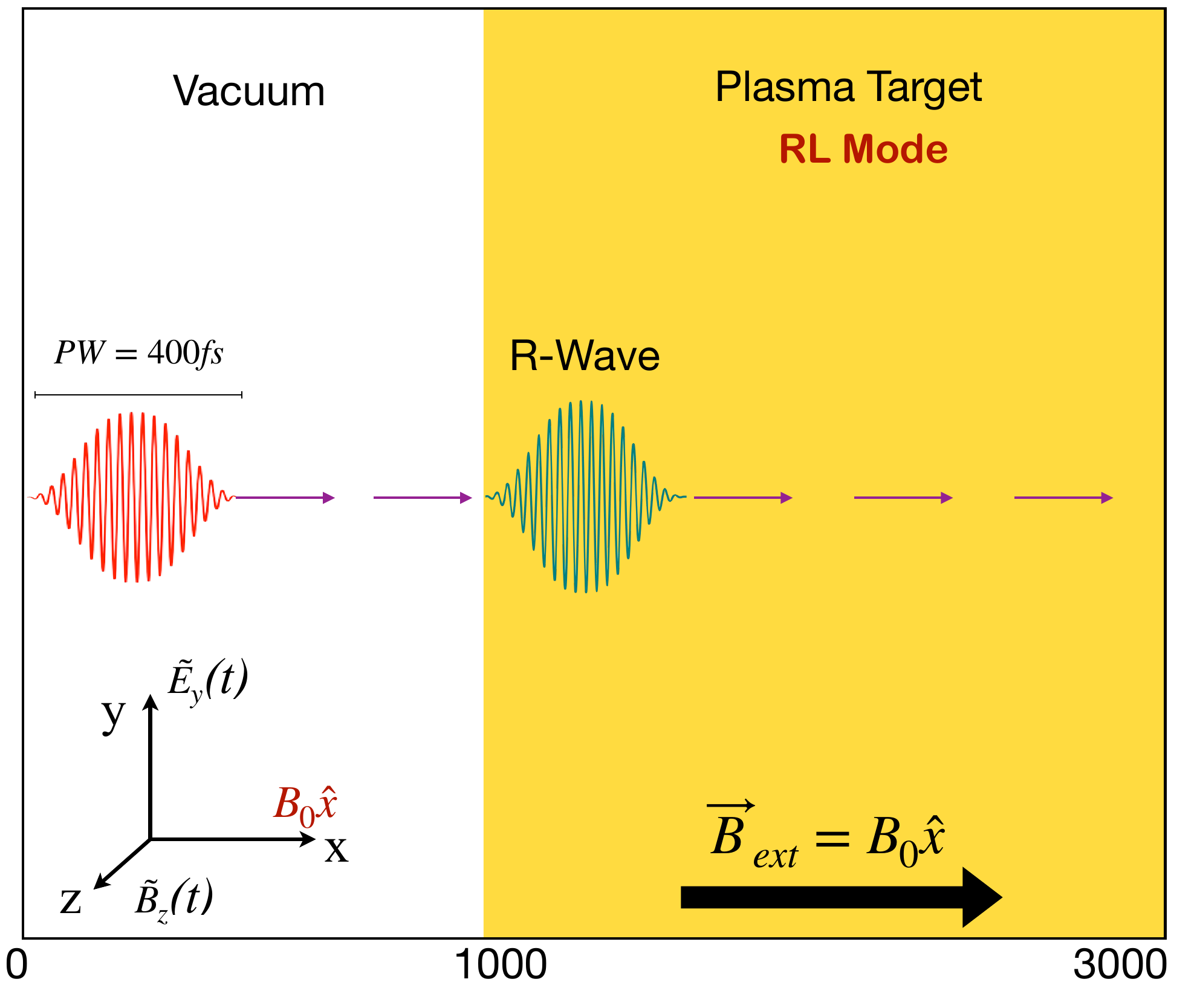}
\caption{\label{schematic2} The schematic configuration for the Electron Cyclotron Resonance (ECR) has been shown.  Here the laser pulse is incident normal to the plasma surface and interacts with magnetized plasma. An external magnetic field along the laser propagation direction is present, as shown in the figure. The incident laser pulse gets partially reflected and partially transmitted inside the plasma.}
\end{figure}

We contrast our studies in section \ref{ECresonance} with a resonant case. In particular, we study the electron cyclotron resonance, the schematic configuration for which has been depicted in 
Fig. \ref{schematic2}. This is a case corresponding to the magnetized plasma resonance. Thus an external magnetic field is applied. The plasma dispersion characteristics in the presence of a magnetic field are quite rich. It permits several pass and stop bands of EM wave frequency even when the plasma is overdense. We choose to work in the frequency regime, which lies in the passband and is close to the electron cyclotron resonance.   

Table - \ref{tab:table1} enlist the laser, plasma, and simulation parameters both in the normalized units as well as their possible values in the standard units that they may correspond to. The OSIRIS4.0 platform uses normalized units.  In our studies, we have depicted the laser parameters for the low-frequency $CO_{2}$ laser. This is so because, for the resonant case that is studied,  the magnitude of the magnetic field for ECR for this laser is in the regime, which is only about an order of magnitude higher than what is experimentally achievable. Thus, one expects that experiments in this regime can be carried out shortly. 

The number of particles per cell is taken to be $16$. 

The time has been normalized by  ${\omega_{pe}}^{-1}$, length by  $ c/\omega_{pe}$ and fields by   $ m_{e}c\omega_{pe}/e$, where $m_{e}$ is the mass of the electron, and $e$ is the charge of the electron. Absorbing boundary conditions have been used for both particles and fields in both directions.

\section{\label{JcrossB} Non-resonant and collisionless  heating }
\begin{figure}
\includegraphics[scale = 0.16]{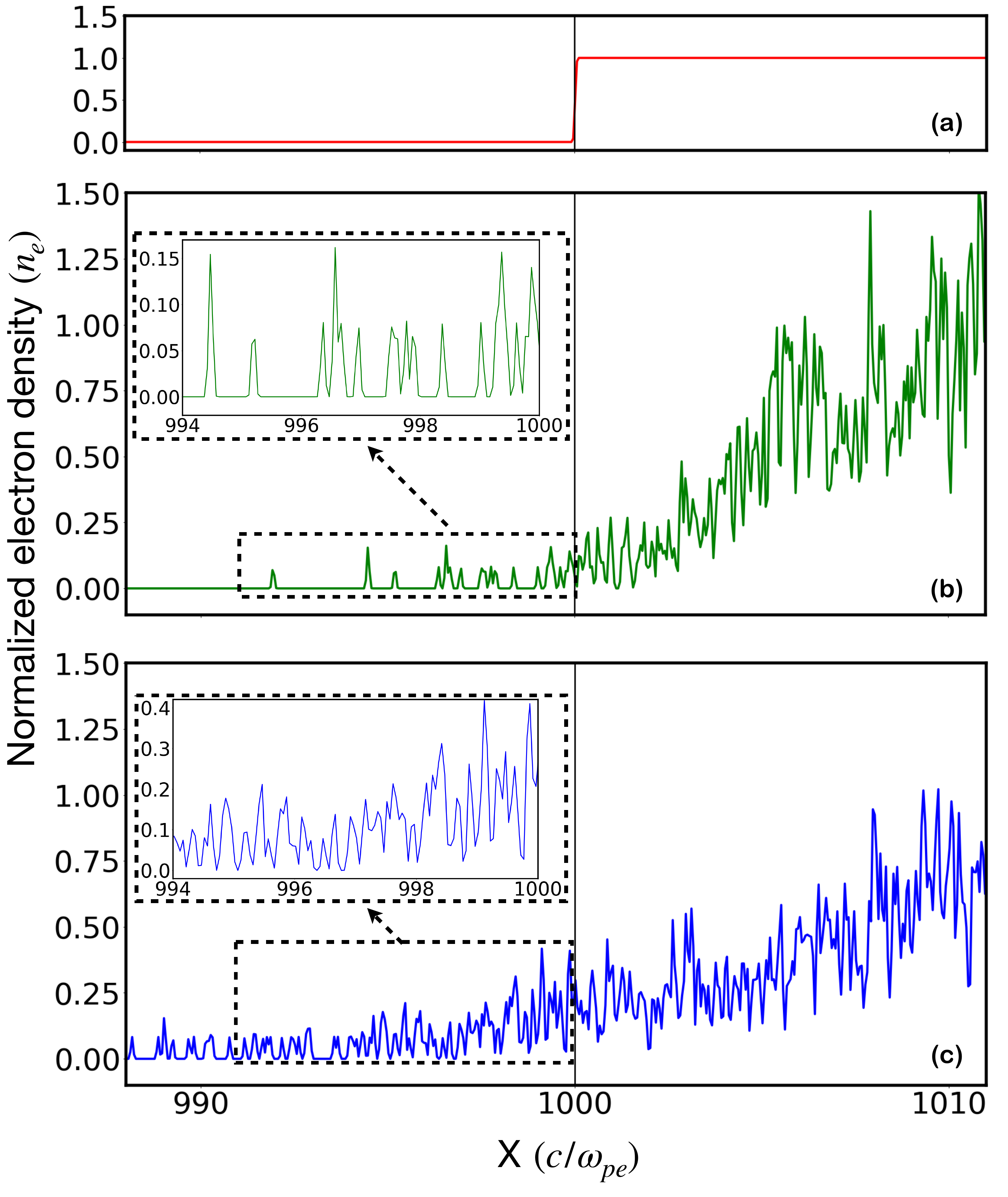}
\caption{\label{elecDens1D} The profile of electron density has been shown at various times. At   $t = 0$, in subplot (a), the initial undisturbed step profile can be seen. The subplots (b) and (c) show the electron density profile after the laser has interacted at t = 1500$\omega_{pe}^{-1}$ and (c) t = 3000$\omega_{pe}^{-1}$ respectively. It can be observed that the electron density is quite smeared up in the vacuum region.}
\end{figure}

We discuss the observations that have been carried out for the two schematic configurations shown in Fig. \ref{schematic}. In both these configurations, the electrons are pulled out from the plasma target into the vacuum, thereby generating sheath electrostatic fields. This can be observed from the plot of Fig. \ref{elecDens1D} where the electron density profile at various times has been shown for the case of  $\vec{J}\times\vec{B}$ configuration. The plasma boundary is at $x = 1000 c/\omega_{pe}$. The subplots (b) and (c) of the figure show the electron density profile at  $t = 1500$ and $3000$ plasma periods, respectively. It is clear that the electrons have been drawn significantly out in the vacuum region.

\begin{figure}
\includegraphics[scale=0.13]{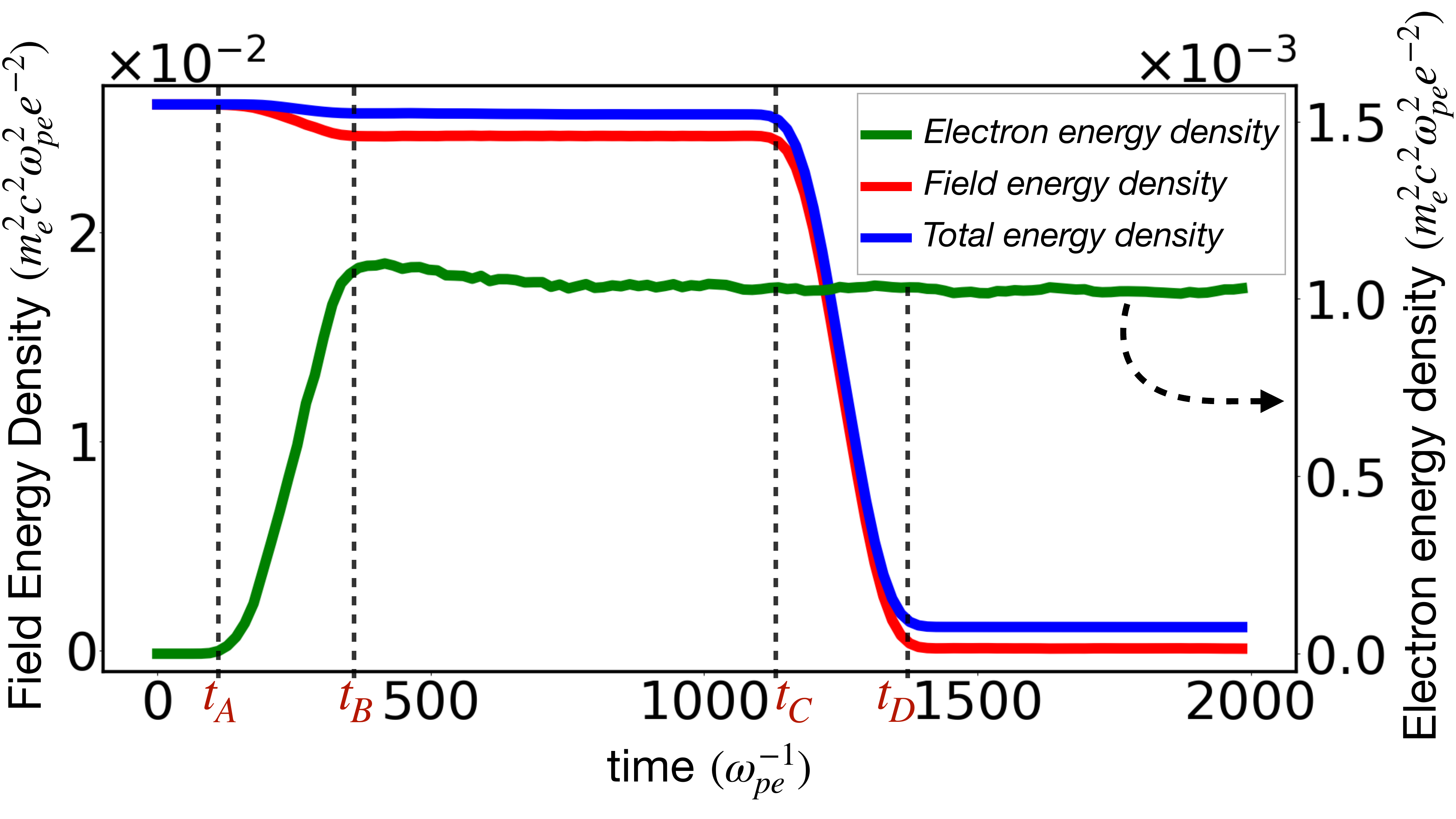}
\caption{\label{energy} Time evolution of the total energy (blue solid line) and the field energy (red solid line) shown by the left $y$ scale. The evolution of the kinetic energy of electrons is depicted by the green solid line by the right $y$ scale. The vertical dashed lines shown at  $t = t_A$ correspond to the time when the laser hits the plasma boundary. The interaction of the laser with the plasma occurs within the time interval $t_b-t_A$. The reflected laser pulse then leaves the simulation box in the interval $t_C$ to  $t_D$. }
\end{figure}

The evolution of total energy density and the field energy density with time has been shown in Fig. \ref{energy}. The left axis shows the scale for these energy densities. The kinetic energy density of the electron is about an order of magnitude smaller and has been depicted by the scale at right in the same figure. The field energy and the total energy are the same initially and remain so till $t = t_A = 98 \omega_{pe}^{-1}$. At this time, the laser hits the plasma boundary. The field energy suffers a small drop from $t = t_A$ to $t=t_B$ while the laser interacts with the plasma surface. The plasma being overdense, the interaction is limited to a surface layer of the order of a few skin depths. It is at this very time interval that the electron kinetic energy,  represented by the green solid line,  increases. The sudden drop in the field energy and the total energy from $t = t_C$ to $t_D$ happens when the reflected laser pulse from the plasma boundary leaves the simulation box. The electron kinetic energy typically remains constant once the laser interaction with the plasma surface ceases at  $t = t_B$. The kinetic energy acquired by the laser, along with the remnant field energy once the laser has left the simulation box, represents an irreversible transfer of energy from the laser to the plasma medium. We now try to understand the underlying process that facilitates this energy transfer process.

\begin{figure}
\includegraphics[scale = 0.14]{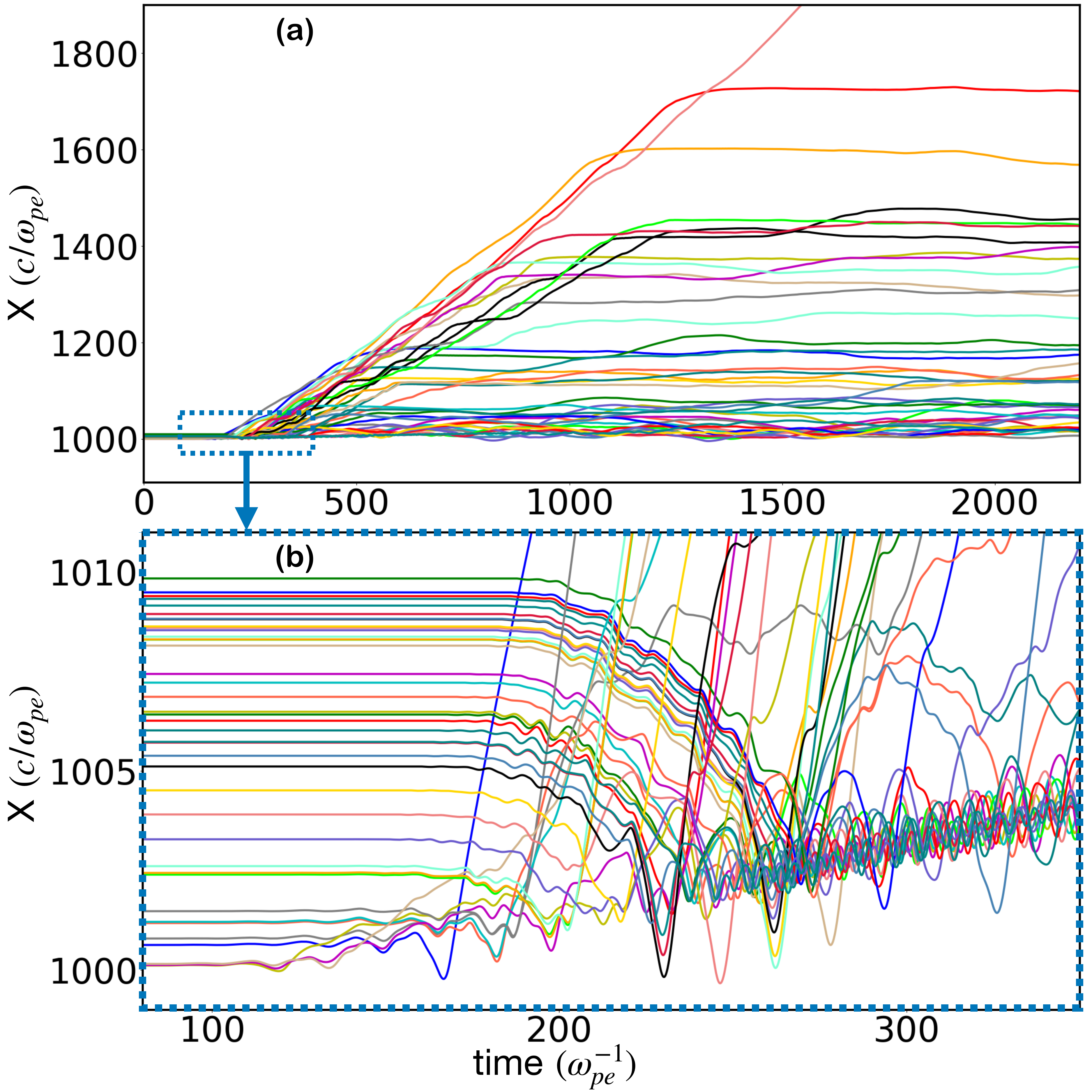}
\caption{\label{trajec} Subplot (a) shows the trajectories of about $40$ electrons which were initially positioned near the plasma surface within the region  $1000$ to $1010$ $c/\omega_{pe}$. The subplot (b) shows the zoomed region. As expected, the electrons extracted out in the vacuum region ( $x< 1000 c/\omega_{pe}$) are also pushed inside the plasma.  }
\end{figure}

The behavior of electron trajectories originating at the plasma surface where the laser plasma interaction have been explored. Trajectories of around $40$ electrons 
within  $10$ skin depths from the plasma surface, i.e., between $1000$ to $1010$ $c/\omega_{pe}$ have been shown as a function of time in  Fig. \ref{trajec}. Many of these electrons have been pulled out in the vacuum region i.e. $x < 1000 c/\omega_{pe}$,  as evidenced by the zoomed subplot (b). The electrons coming out are also observed to return inside the target once the laser cycle changes.  From the slope of the trajectories in this figure, the particle speed can be estimated, and it is clear that often their speed while returning is much higher than with what they left.  Thus there is an overall gain in energy by the particles during the process as they get pulled out and then pushed back again in the plasma target. 

\begin{figure}
\includegraphics[scale = 0.12]{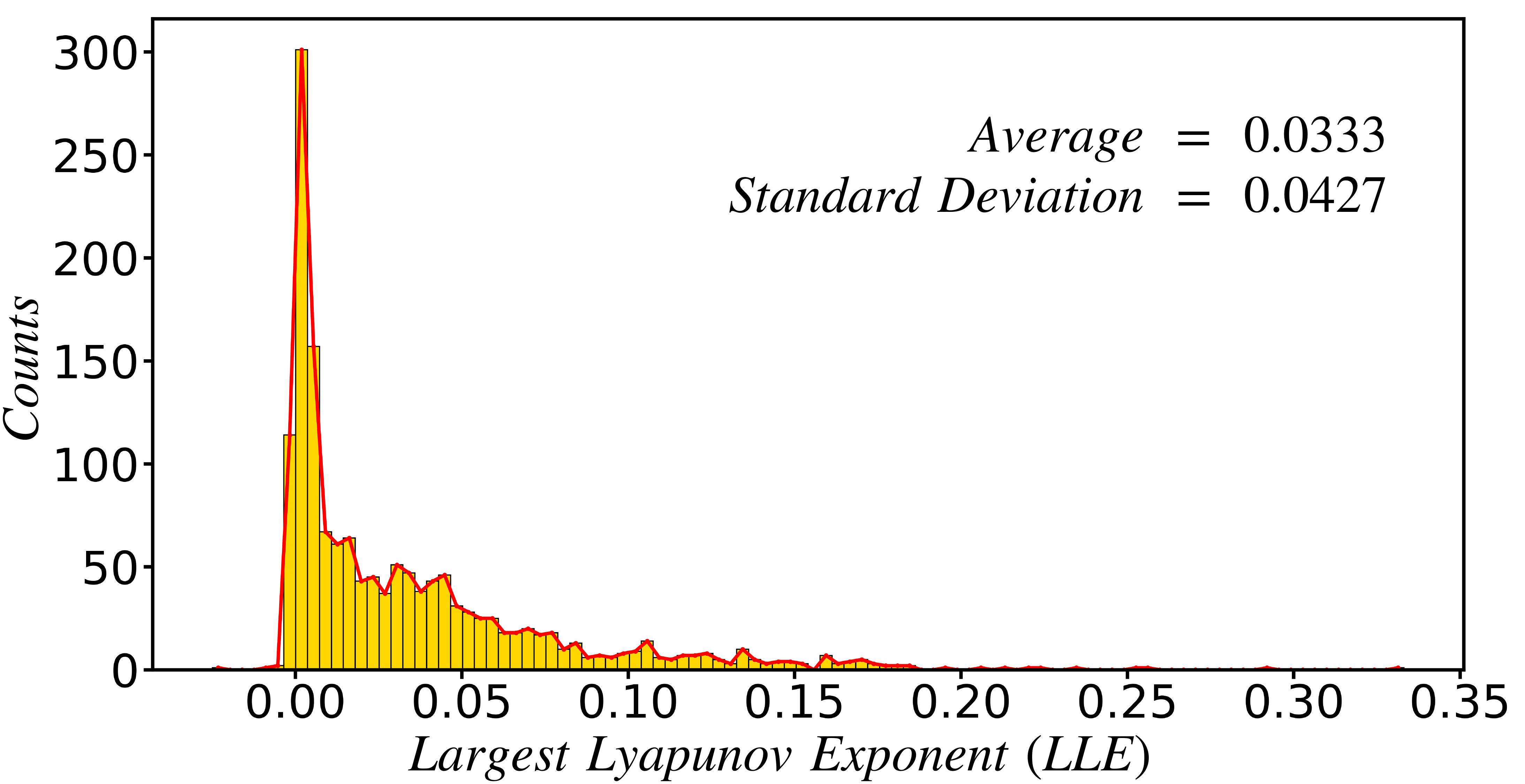}
\caption{\label{LLE} The distribution of the Largest Lyapunov Exponent (LLE) for a set of $1500$  electrons (initial position within  $1000$ to $1010$ $c/\omega_{pe}$) has been shown. This figure is for the case where  $\vec{J}\times\vec{B}$ is the operative mechanism for energy absorption.  }
\end{figure}

We now explore the behavior of the electron trajectories quantitatively to see whether there are chaotic traits in these trajectories. About  $1500$ electron trajectories originating near the plasma surface from $1000$ to $1010$ $c/\omega_{pe}$ were analyzed while the laser pulse was interacting with the plasma medium. Thereafter, their  Largest Lyapunov Exponent (LLE) 
\cite{rosenstein1993practical} was determined. The distribution of the Largest Lyapunov Exponent (LLE) has been shown in Fig. \ref{LLE}.  It is evident from the figure that most of the particles have a positive value of LLE.  The average value of LLE in this case is found to be about $0.0333$, which is positive. The observed standard deviation is $0.0427$, which indicates a significant spread in the value of LLE. This confirms that the electron trajectories are chaotic in the combined laser and the self-consistent field. The chaotic nature of the electron trajectories may facilitate the energy absorption process in this case by providing the appropriate phase shift between the velocity and the laser electric field.

\begin{figure}
\includegraphics[scale = 0.12]{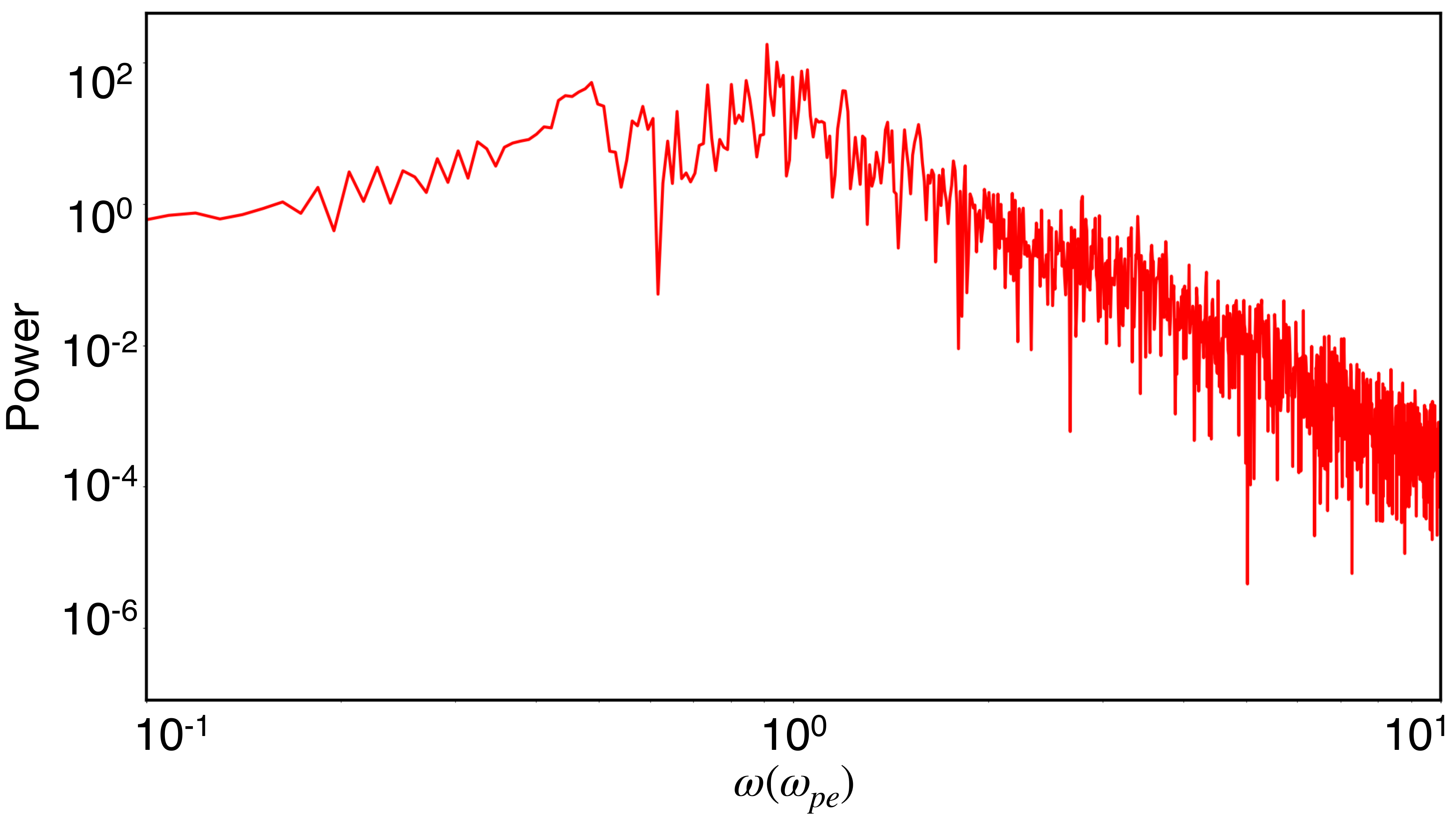}
\caption{\label{FFT1D} Power spectrum for time series of $x$-component of velocity ($\mathrm{v}_x$) has been shown for the case of $\vec{J} \times \vec{B}$.}
\end{figure}

\begin{figure}
\includegraphics[scale = 0.13]{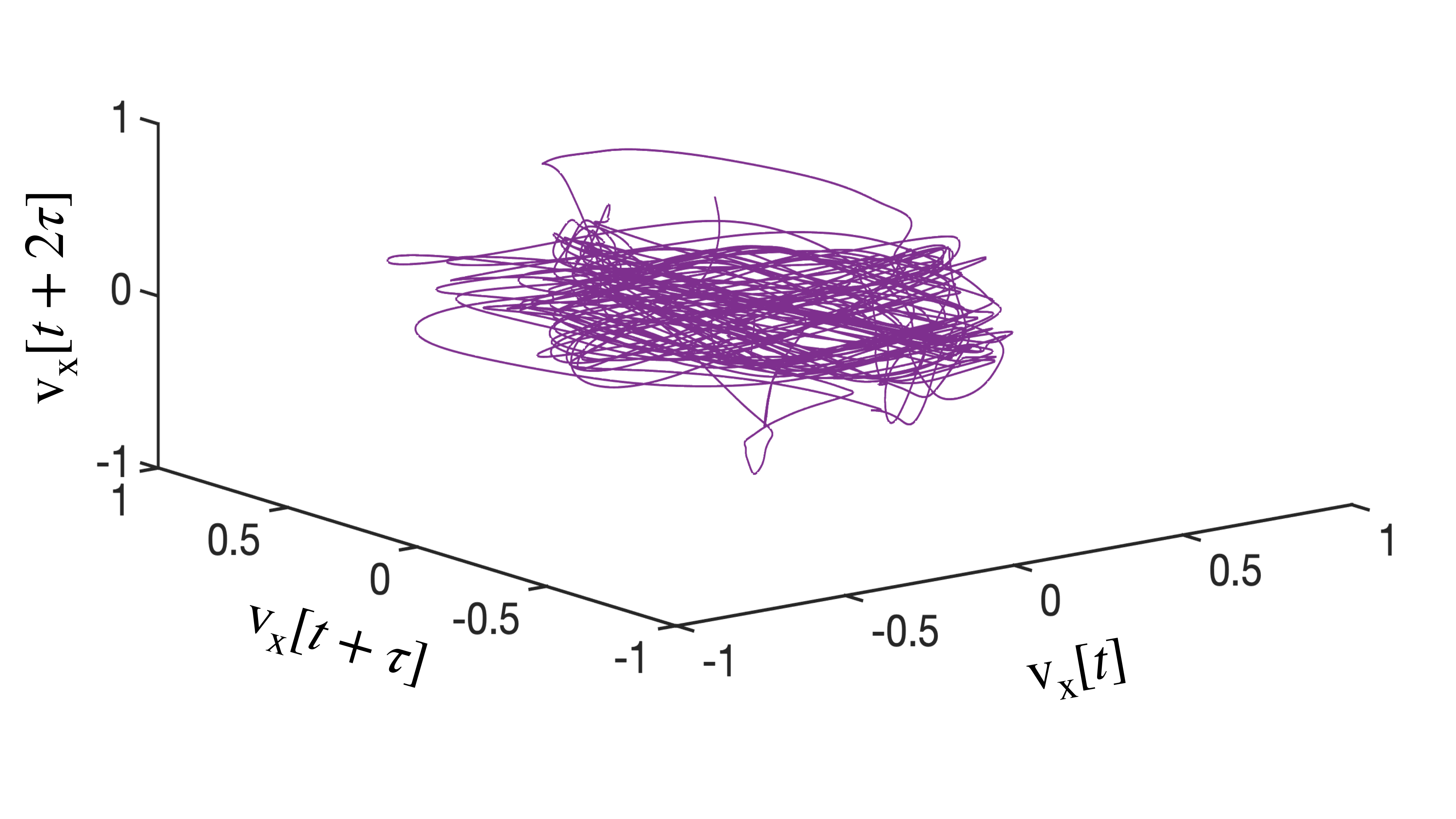}
\caption{\label{reconstruct1D} A 3-D  phase space reconstruction has been shown for the attractor with the time series data of the  $x$-component of velocity ($\mathrm{v}_x$). This is for the case where the $\vec{J} \times \vec{B}$ mechanism is operative.}
\end{figure}

\begin{figure}
\includegraphics[scale = 0.13]{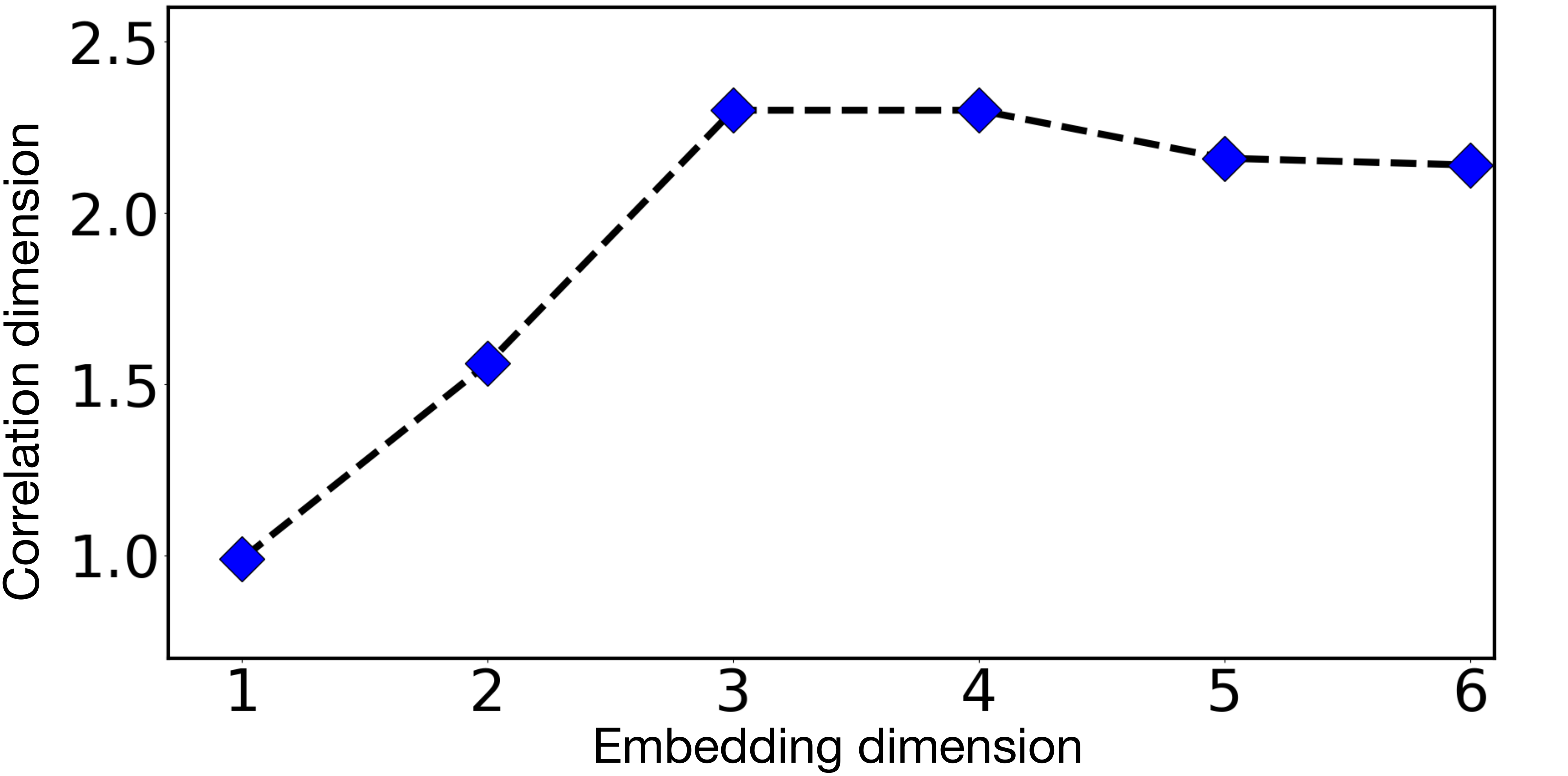}
\caption{\label{embedd1D}  A plot of correlation dimension versus embedding dimension for the case where  $\vec{J} \times \vec{B}$ mechanism is operative.}
\end{figure}

We now analyze some other aspects of the time series data of the velocity field ${\mathrm{v_x}}$. The power spectrum shown in  Fig. \ref{FFT1D} is considerably broad. The phase space reconstruction and the determination of the correlation dimension has also been carried out \cite{baker1996chaotic}. 
A reconstructed 3-D attractor for the time series of ${\mathrm{v_x}}$ is shown in Fig. \ref{reconstruct1D}. To resolve the system's structure in the reconstructed phase space, the minimum embedding dimension $m$ has been determined, and it is found to be $3$ through the “false nearest neighbor” method \cite{kennel1992determining}. With $m = 3$ chosen, no self-intersection was observed. We then evaluate the correlation dimension \cite{grassberger1983characterization} ($d$). It should be independent of embedding dimension $m$, which gives information about the attractor that is effectively embedded in a higher dimensional space. Our analysis found a correlation dimension of $d = 2.3$. It should be noted that by increasing the embedding dimension beyond $3$, the correlation dimension becomes independent of $m$, as shown in Fig. \ref{embedd1D}. In fact this feature helps distinguish the observed dynamics from stochastic to that exhibiting chaos. For the latter, the attractor is strange with a non-integral dimension. We have carried out a similar analysis for some other particles also and observed that all the particles near the surface have chaotic attributes. Though the quantitative values of the   Lyapunov index differ slightly, they are all positive.  
 
The second case that we consider is that of vacuum heating, for which the configuration is shown in the schematic of Fig. \ref{schematic}(b). A  p-polarised laser is incident obliquely on an overdense plasma target. The electric field component normal to the vacuum plasma interface pulls and pushes the electrons from the plasma into the vacuum and back, thereby generating sheath electric fields. The oblique incidence requires a  2-D spatial geometry for simulation for studying this configuration.

\begin{figure}
\includegraphics[scale=0.13]{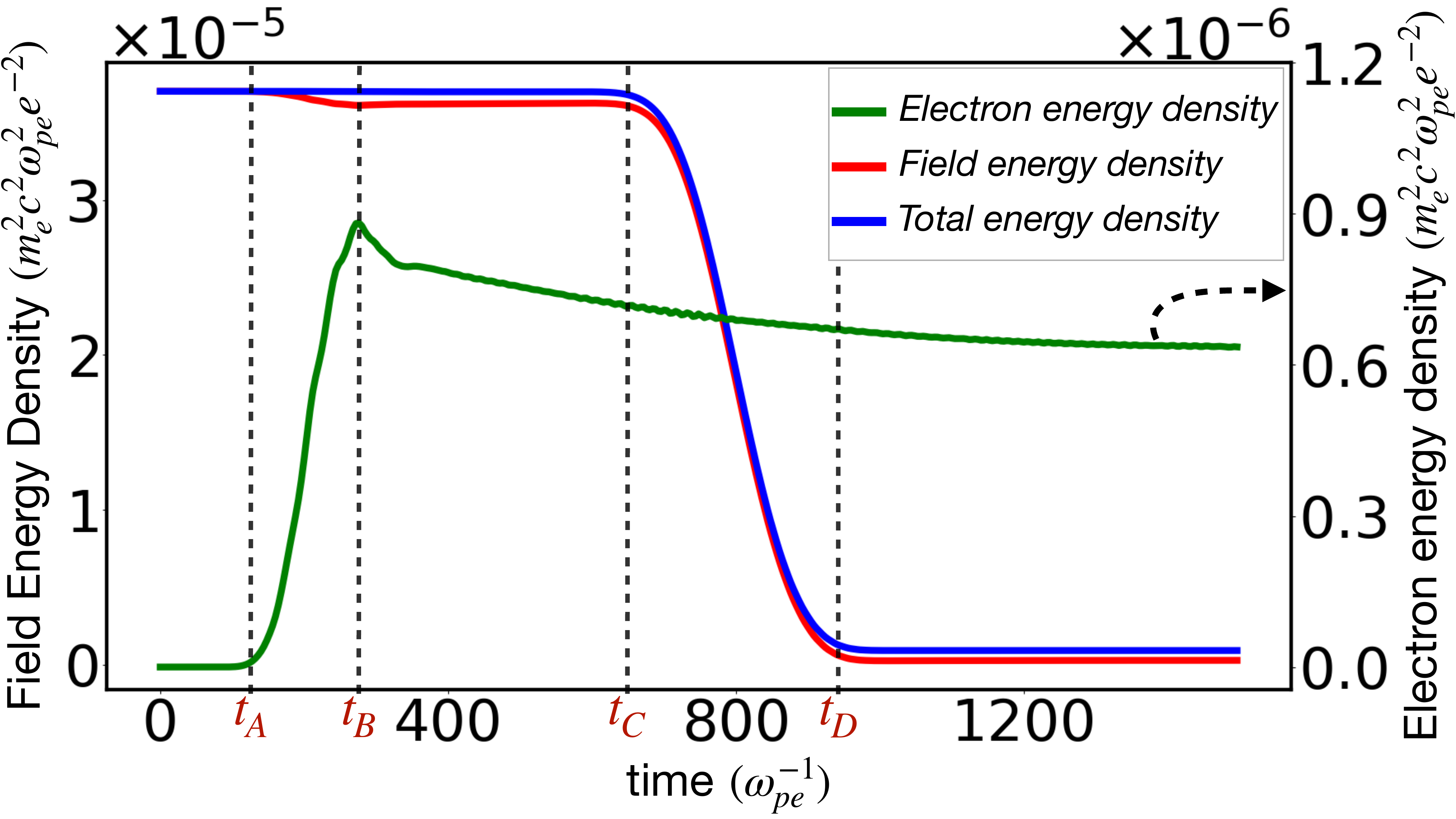}
\caption{\label{energy2D} Time evolution of field energy density and kinetic energy density of electrons for vacuum heating process. The time stamps denoted by the vertical lines at $t_A, t_B, t_C,$ and $t_D$ are the same as described in Fig. \ref{energy}. }
\end{figure}

\begin{figure}
\includegraphics[scale = 0.12]{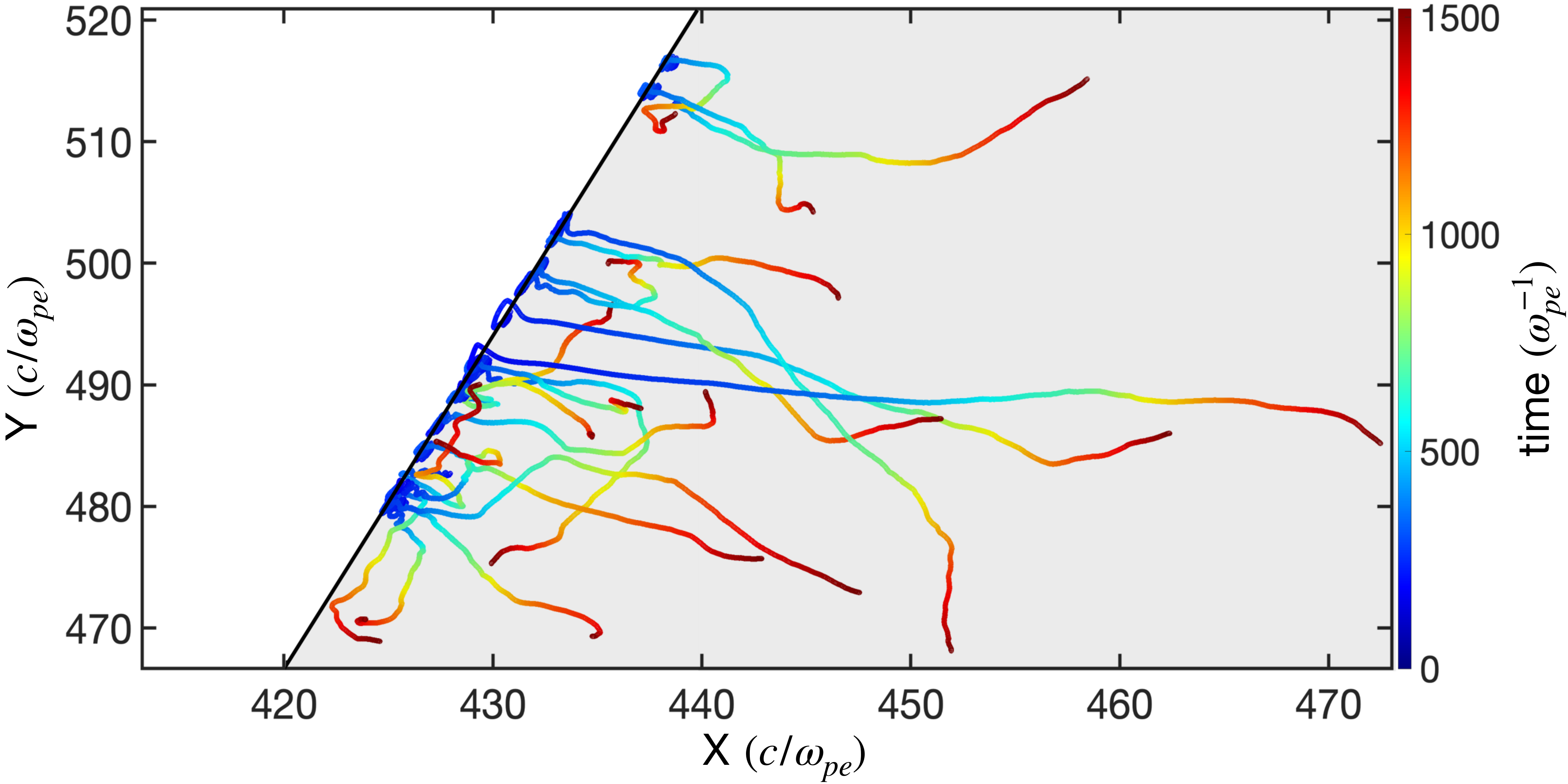}
\caption{\label{trajec2D} Trajectories of $20$ electrons initially positioned near the plasma surface in the case of vacuum heating shown in the X-Y plane.}
\end{figure}

\begin{figure}
\includegraphics[scale = 0.12]{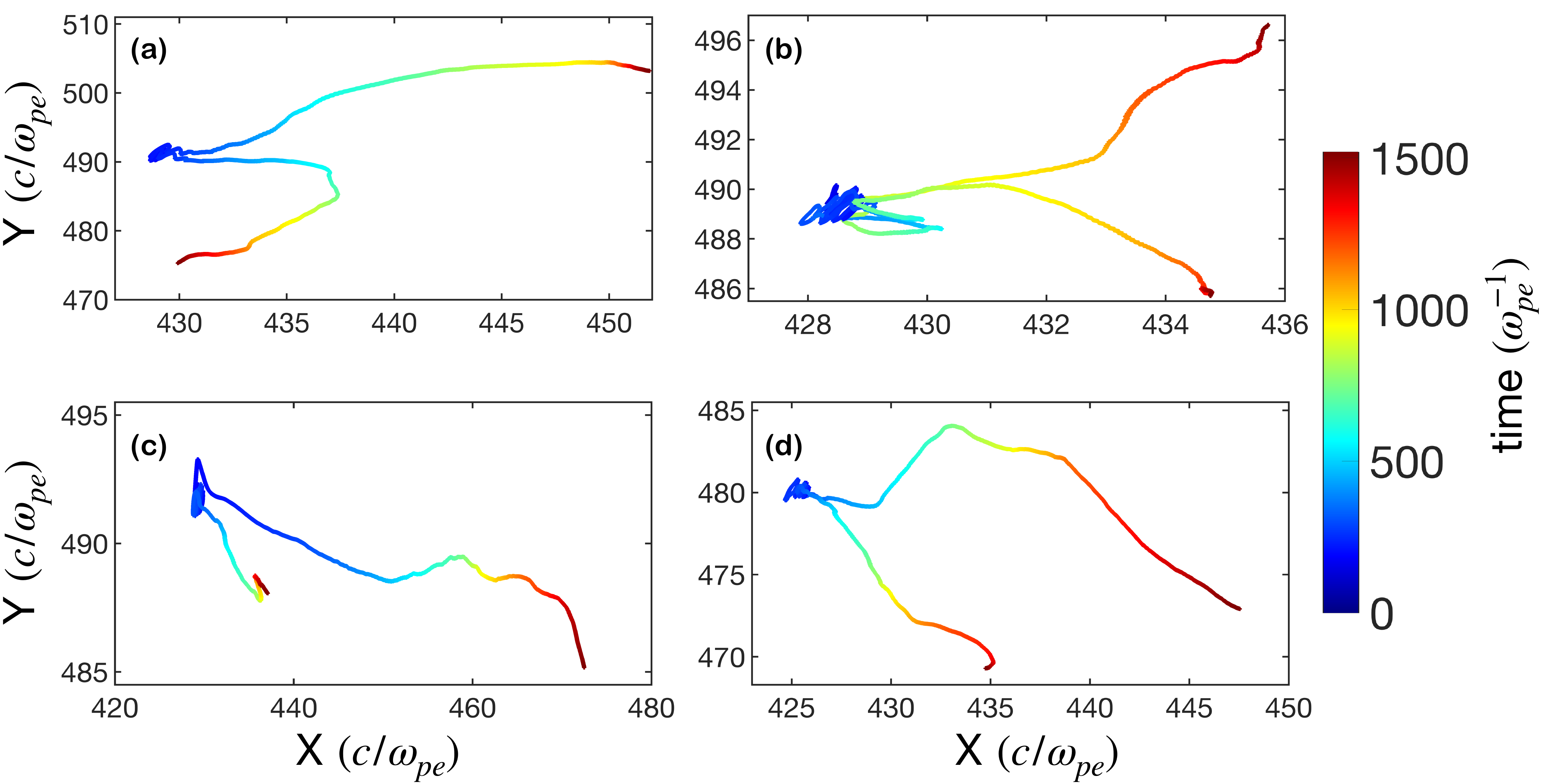}
\caption{\label{fourTraj2D} The divergent trajectories of four pairs of initially closely placed electrons for the case of vacuum heating have been shown.}
\end{figure}

\begin{figure}
\includegraphics[scale = 0.12]{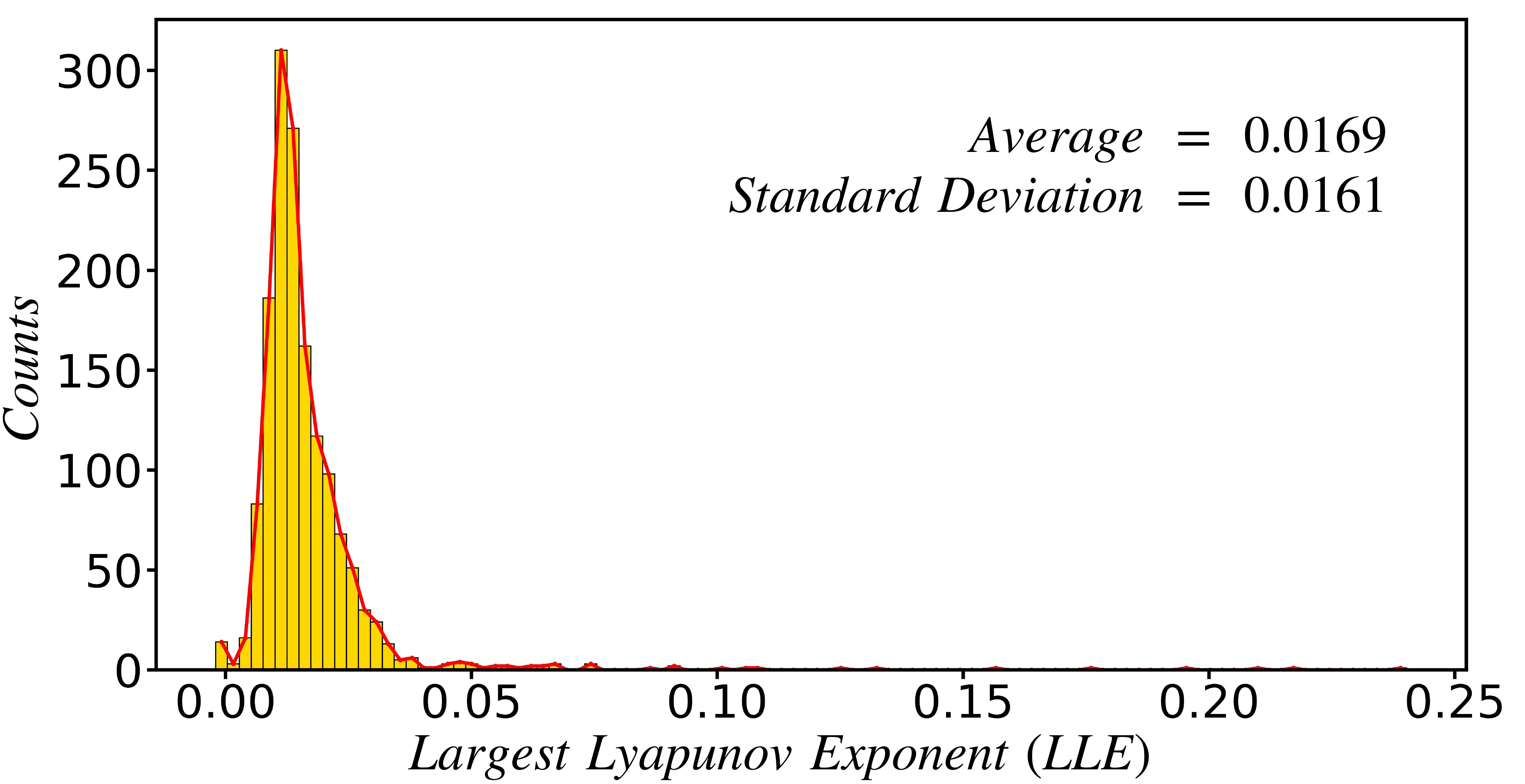}
\caption{\label{LLE_2D} The distribution of the Largest Lyapunov Exponent (LLE) for  $1500$ electrons has been shown when the vacuum heating mechanism is operative.  The average value of LLE as well as its standard deviation in this case, is less than the case of $\vec{J} \times \vec{B}$ heating shown in Fig. \ref{LLE}.}
\end{figure}

\begin{figure}
\includegraphics[scale = 0.12]{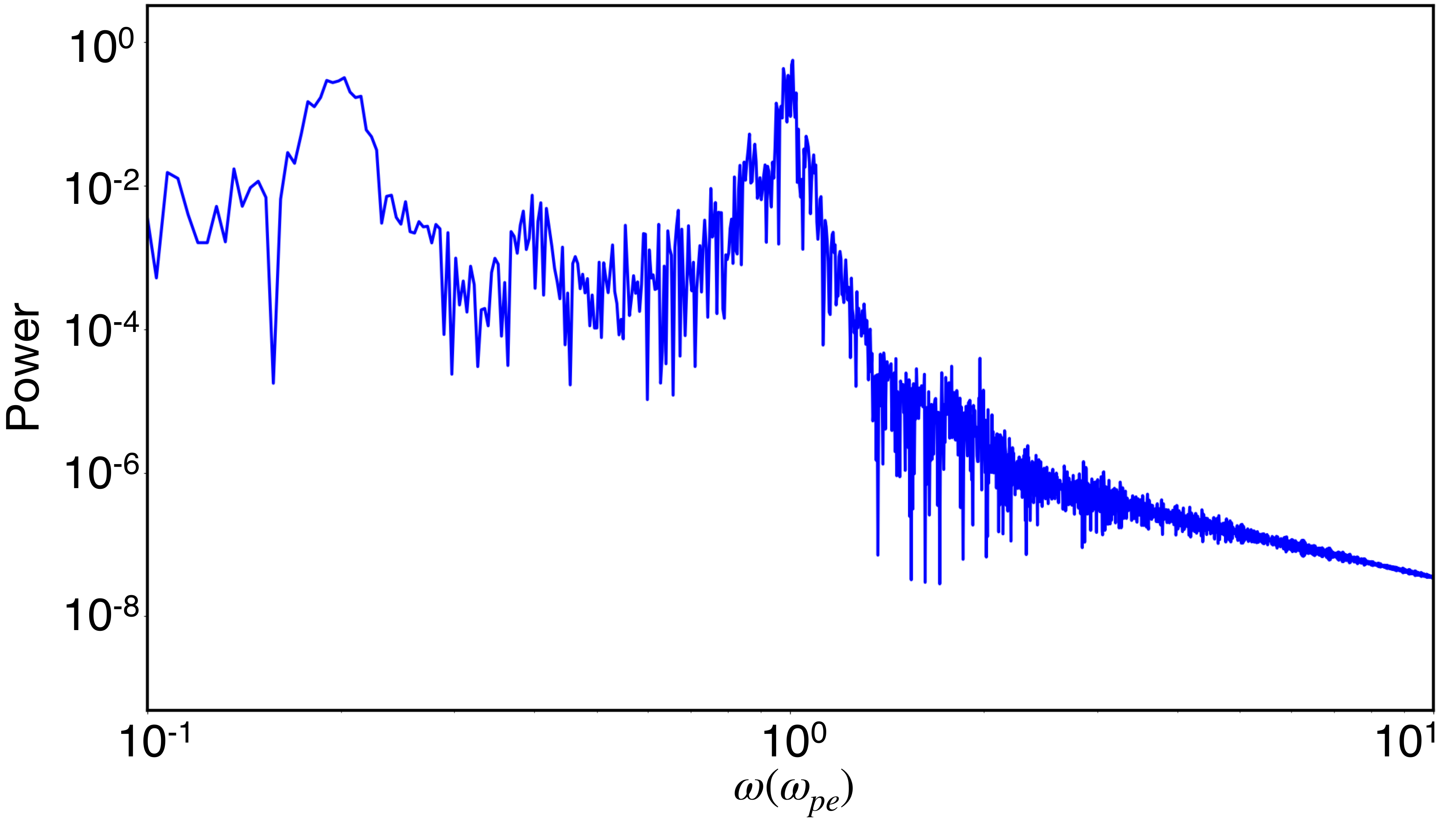}
\caption{\label{FFT2D} Power spectrum for time series of $\mathrm{v}_x$ in case of vacuum heating. The broad spectrum with a peak at the plasma oscillation frequency can be observed.}
\end{figure}

\begin{figure}
\includegraphics[scale = 0.13]{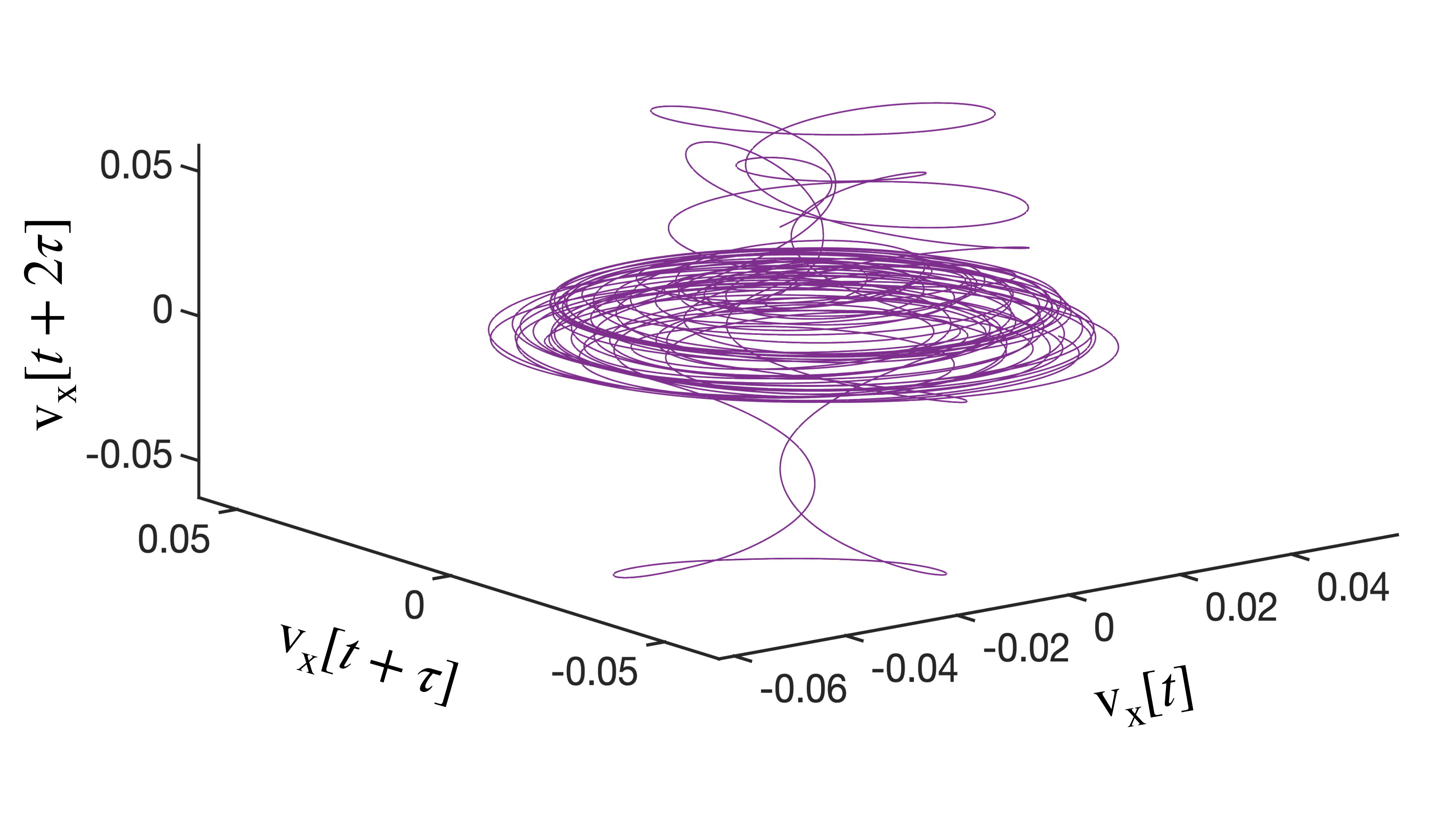}
\caption{\label{reconstruct2D} The 3-D reconstructed phase space attractor using the time series data of the $x$-component of the velocity  $\mathrm{v}_x$ has been shown when the vacuum heating mechanism is operative.}
\end{figure}

\begin{figure}
\includegraphics[scale = 0.13]{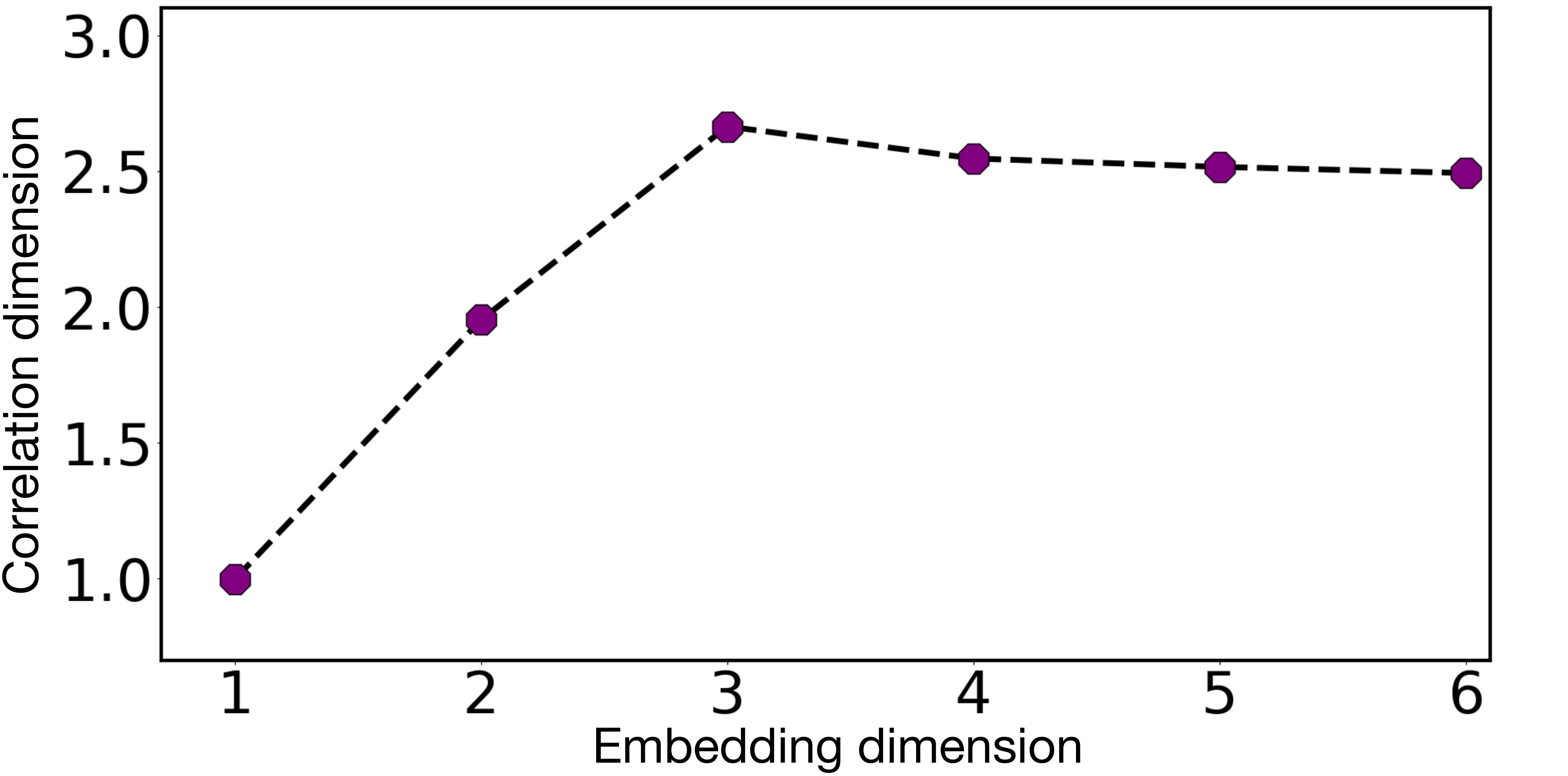}
\caption{\label{embedd2D} Correlation dimension versus Embedding dimension has been shown when the vacuum heating mechanism is operative.}
\end{figure}
The evolution of field energy density and the kinetic energy density of electrons have been depicted in Fig. \ref{energy2D} by various curves. Just as described in Fig. \ref{energy} for the 
$\vec{J} \times \vec{B}$ case,  here too the vertical dashed lines denoting  $t_A, t_B $, $t_C$ and $t_D$ represent various times for which the following happens. Here $t_A = 150 \omega_{pe}^{-1}$ is the time when the laser hits the plasma boundary, the laser-plasma interaction continues till $t_B$, and from $t_C$ to $t_D$ the reflected laser pulse leaves the simulation box.  The red solid line represents the field energy, and the blue curve is the total energy. 
In this case also, a certain fraction of laser energy is retained in the plasma medium after the laser pulse has left the simulation box. This is both in the form of the kinetic energy of the electrons as well as field energy. It is worth noting that the percentage of energy transferred to electrons in this particular case is smaller than that of the earlier case. Here, it is about  $1 \%$ of the total energy density that is retained as the kinetic energy in the electrons, whereas in the former case described by Fig. \ref{energy}, it was about $3 \%$. 

We now explore the behavior of the electron trajectories picked up on the plasma surface for this particular case. The trajectories of about  $20 $ electrons have been shown in Fig. \ref{trajec2D}. It can be observed from Fig. \ref{fourTraj2D} that trajectories of even those electrons that originate from locations nearby significantly diverge. 
The distribution of the Largest Lyapunov Exponent (LLE) for $1500$ electrons has been plotted in Fig. \ref{LLE_2D}. It is observed that the LLE for all the electrons, though distinct from each other, are positive. The average and standard deviation of the LLE in this case are $0.0169$ and $0.0161$, respectively. 
This is comparatively smaller than the previous $\vec{J} \times \vec{B}$ case we considered earlier. 

The time series of the ${\mathrm{v_x}}$ component of velocity has again been analyzed. The power spectrum of ${\mathrm{v_x}}$ is found to be broad (Fig. \ref{FFT2D}) and shows a peak at the plasma frequency, unlike the previous case. The presence of a peak riding on top of the broad spectrum suggests that while the electrons are driven in and out at the laser frequency, they excite the characteristics plasma oscillations in the medium. 
 
The phase reconstruction has also been carried out (Fig. \ref{reconstruct2D}). The 
correlation dimension of the attractor was found to be typically around $2.6$ (Fig. \ref{embedd2D}) for this particular case.

It is interesting to note that while both the cases of $\vec{J} \times \vec{B}$ and vacuum heating exhibit the chaotic nature of surface electron trajectories, the former case has a higher value of the LLE compared to the latter. The energy fraction that gets absorbed is also found to be higher in the case for which LLE is higher.  Such a trend is another indicator that the chaotic nature of the trajectories has a role in determining energy absorption.

In the next section, we carry out a simulation and similar analysis for a resonant case. 
 
\section{\label{ECresonance} Resonant case}
We now contrast the above schemes of laser energy absorption with a purely resonant case in which no electrostatic field gets generated.  We rule out the  $\vec{J} \times \vec{B}$ and vacuum heating mechanism by choosing the laser intensity to be non-relativistic and making it fall on the target at normal incidence. The plasma target is immersed in an external magnetic field, as shown in Fig. \ref{schematic2}. The magnetic field is chosen to be strong enough to satisfy the magnetized dispersion relation \cite{boyd2003physics}.  Despite the plasma being overdense there are several pass and stop bands permissible for the laser in this regime. The regime of laser interacting with the magnetized plasma has also been of considerable interest lately and has been pursued in several contexts (e.g. energy absorption, harmonic generation etc., \cite{maity2022mode,dhalia2023harmonic,juneja2024enhanced, goswami2022observations,vashistha2020new,mandal2021electromagnetic,dhalia2024absorption}. There are a variety of magnetized plasma modes with which the laser/EM wave frequency might resonate.  We choose to study the case of electron cyclotron resonance here.  We investigate the energy transfer and the behavior of particle trajectories in this case and contrast it with the non-resonant cases presented in section \ref{JcrossB}.

\begin{figure}
\includegraphics[scale=0.13]{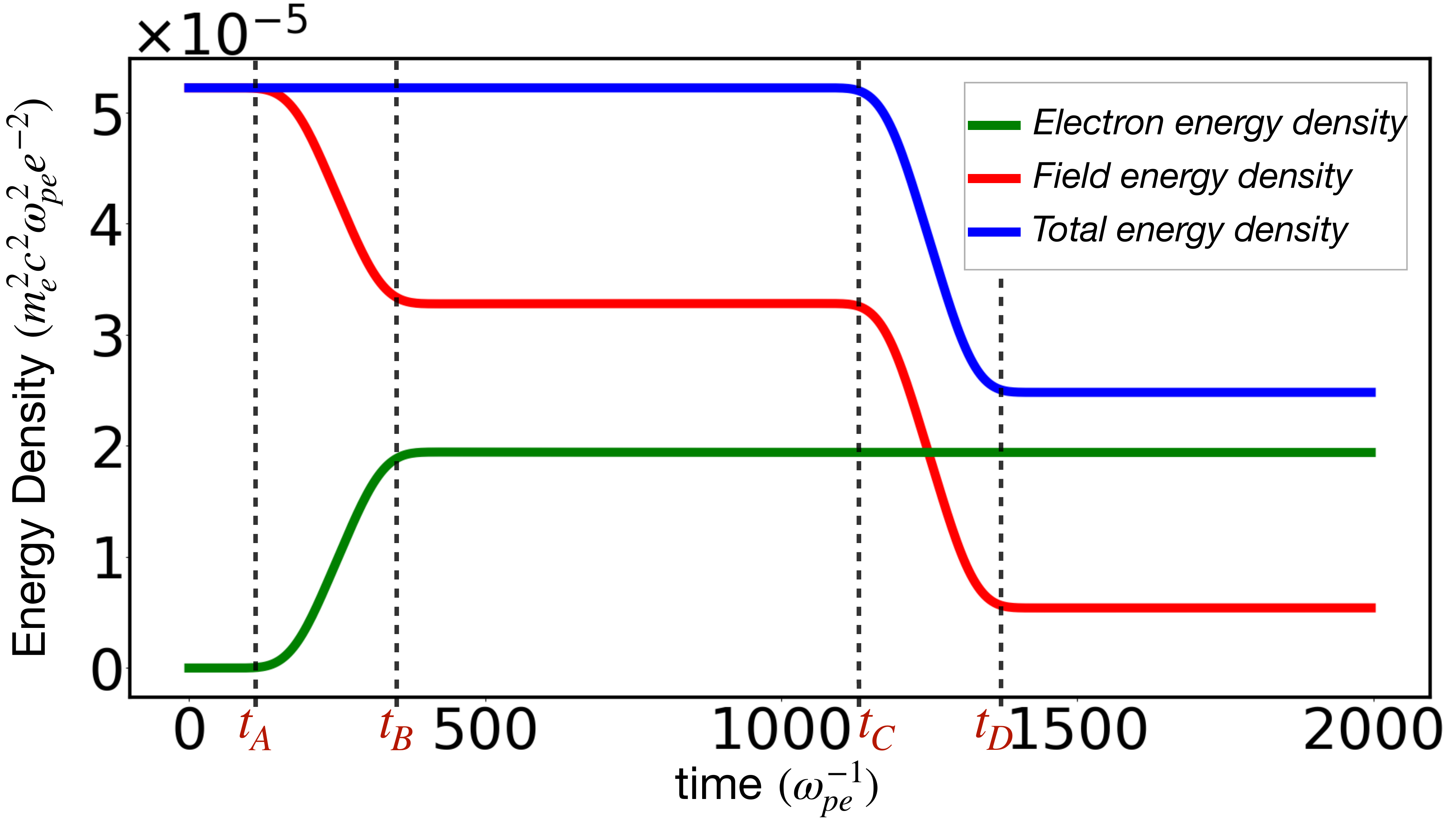}
\caption{\label{MagnetizedEnergy} Time evolution of total energy (blue solid line), field energy(red solid line), and electron kinetic energy (green solid line)  has been shown for the ECR resonance studies. The kinetic energy acquired by electrons is quite significant compared to the other non-resonant cases. }
\end{figure}

The frequency of the laser field is chosen to be $0.31\times10^{15} Hz$. This lies in the $R$ mode pass band of the magnetized plasma dispersion relation \cite{boyd2003physics} for  $B_0 = 0.4$ in normalised units. For this value of the external magnetic field, the laser frequency is close to the electron cyclotron resonance.  This choice in fact conforms with the inference drawn in the study carried out by  Juneja {\it{et al.}} \cite{Juneja_2023} for optimum energy absorption near resonance,  despite the absence of non-resonant processes of electrostatic field generation at work. This can be seen from the plot of  Fig. \ref{MagnetizedEnergy}, which illustrates the evolution of various kinds of energies. Here, as the laser pulse touches the surface of the plasma target  $t_A$ it gets partially reflected and partially transmitted from the plasma surface.    
The interaction with the incident pulse continues till $t_B$. 
At $t_C $, the reflected laser pulse starts leaving the simulation box, and by  $t_D$, it is completely out of the simulation box. It should be observed that the kinetic energy of the electrons increases during the time interval $t_B-t_A$. It is interesting to note that for this particular case, the percentage energy acquired by the electrons is very significant,  almost $\sim 37 \%$.

\begin{figure}
\includegraphics[scale = 0.12]{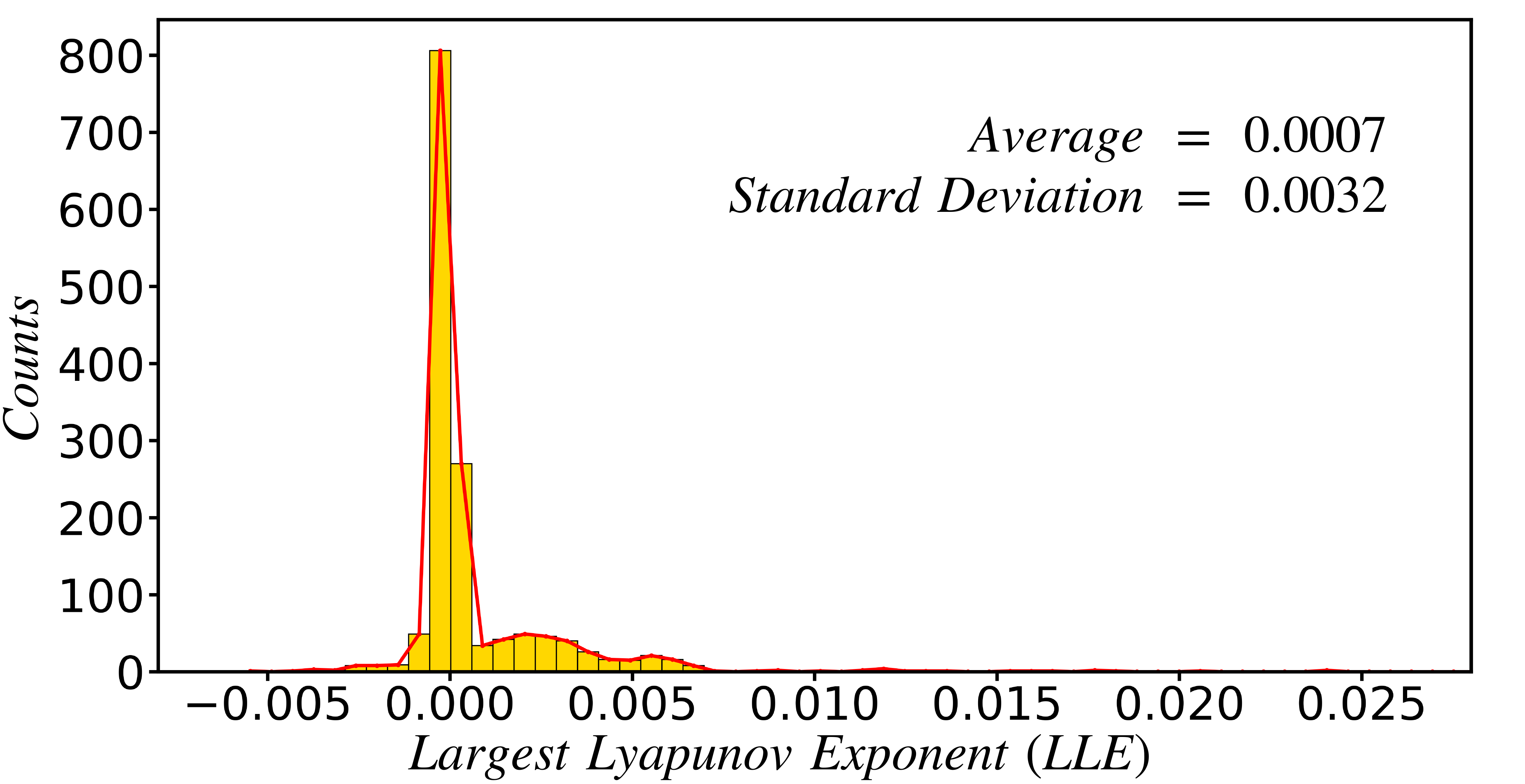}
\caption{\label{LLE_RL_surface} Distribution of the Largest Lyapunov Exponent (LLE) for  $1500$ surface particles (originally lying within $1000$ to $1010$ $c/\omega_{pe}$) in  ECR resonance. The LLE has very small values in this case compared to others.}
\end{figure}

We now repeat the analysis for understanding the behaviour of electron trajectories which have acquired significant kinetic energy in this particular case.  Again a set of about $1500$ electron particles are chosen from the surface plasma region extending from $1000$ to $1010 c/\omega_{pe}$. The distribution of the  Largest Lyapunov Exponents (LLE)  is plotted in Fig. \ref{LLE_RL_surface}. The values of LLE for most of the particles are observed to be very close to zero. In fact, for many cases, even a negative value of LLE has been observed. The average value of LLE is about $0.0007$ with a standard deviation of $0.0032$. 
  
One thus observes that even though this case provides for a very high energy absorption percentage, it is governed by resonant interaction, and chaos plays a negligible role here.

\begin{figure}
\includegraphics[scale = 0.125]{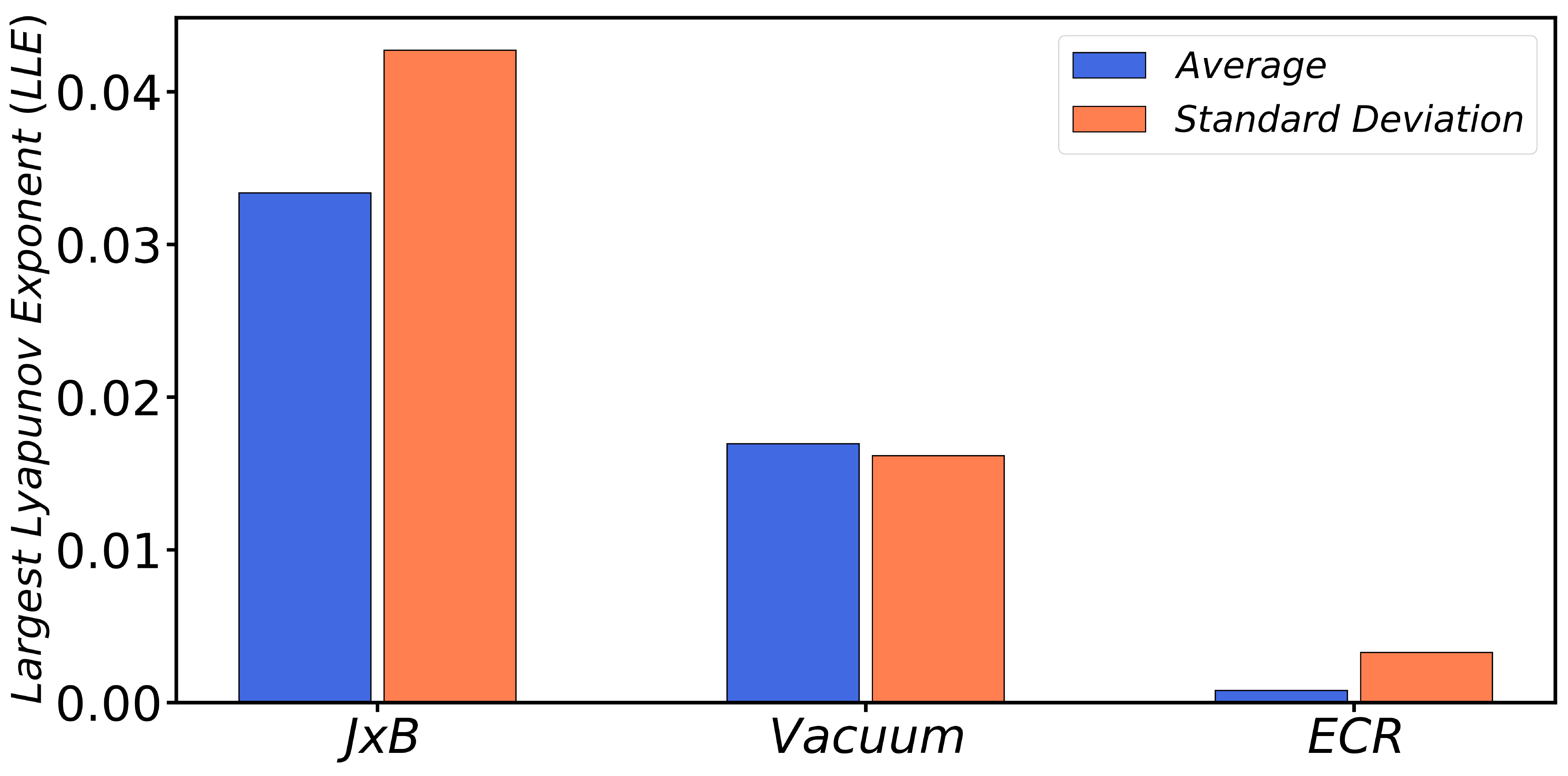}
\caption{\label{LLE_All} A comparison of the average and standard deviation of the Largest Lyapunov Exponent (LLE) for all three cases. The  Lyapunov index has noticeably high values when the absorption process is mediated through the non-resonant mechanism. Furthermore, it is higher in the first case of   $\vec{J} \times \vec{B}$ heating compared to vacuum heating. The energy absorption follows the same trend as the LLE. In contrast, though the energy absorption is highest for the ECR resonance, the LLE is found to be the smallest.}
\end{figure}

\begin{figure}
\includegraphics[scale = 0.17]{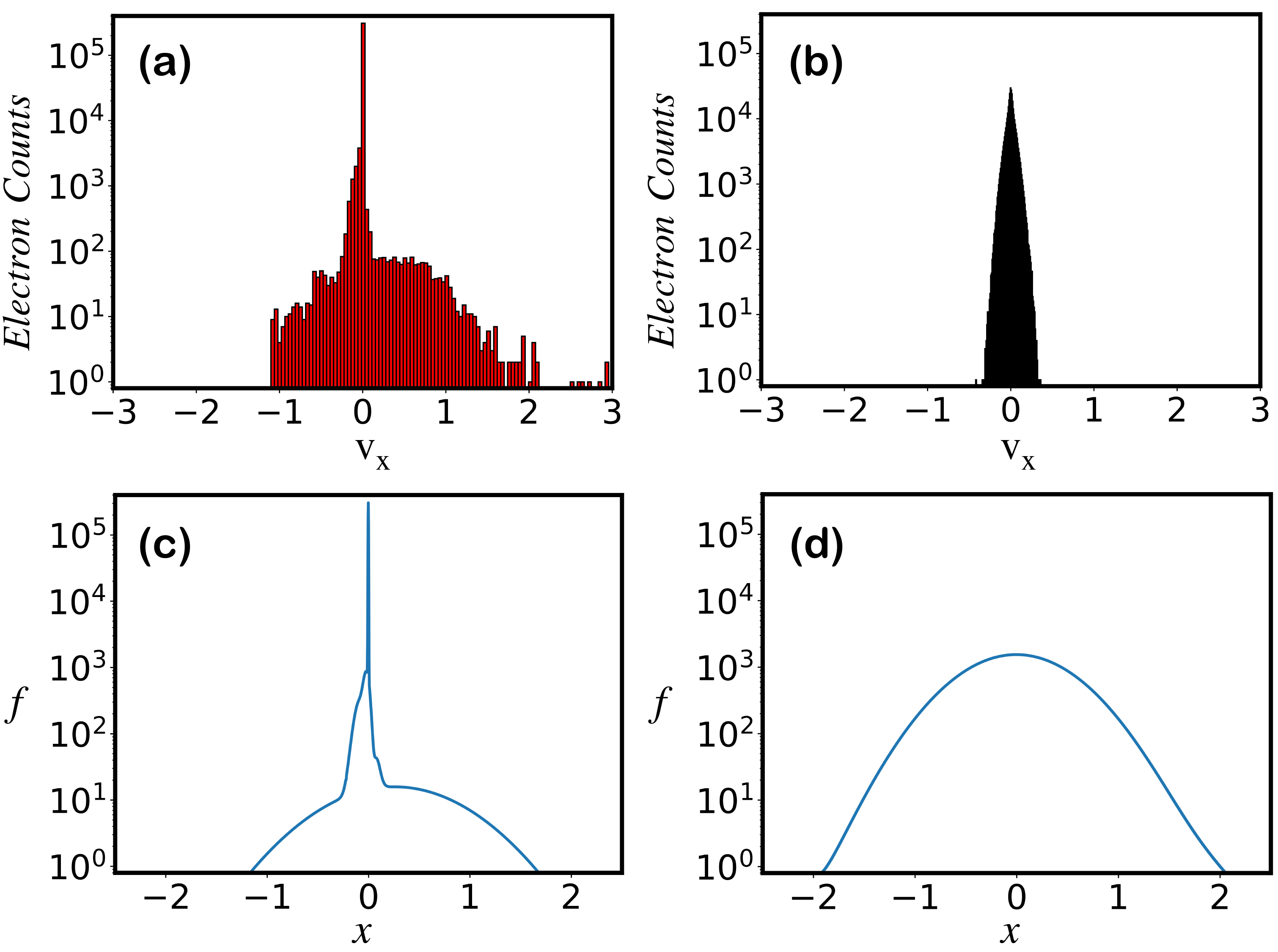}
\caption{\label{Diffusion_JxB} Diffusion in case of $\vec{J}\times\vec{B}$ heating. Snapshots (a) and (b) show the $\mathrm{v_x}$ component of velocity distributions at times  $t_1 = 320$  and  $t_2 = 3500$, respectively, from PIC simulations. Diffusion of initial distribution profile to single maxwellian from fluid simulations have been shown in snapshots (c) and (d).}
\end{figure}

\begin{figure}
\includegraphics[scale = 0.17]{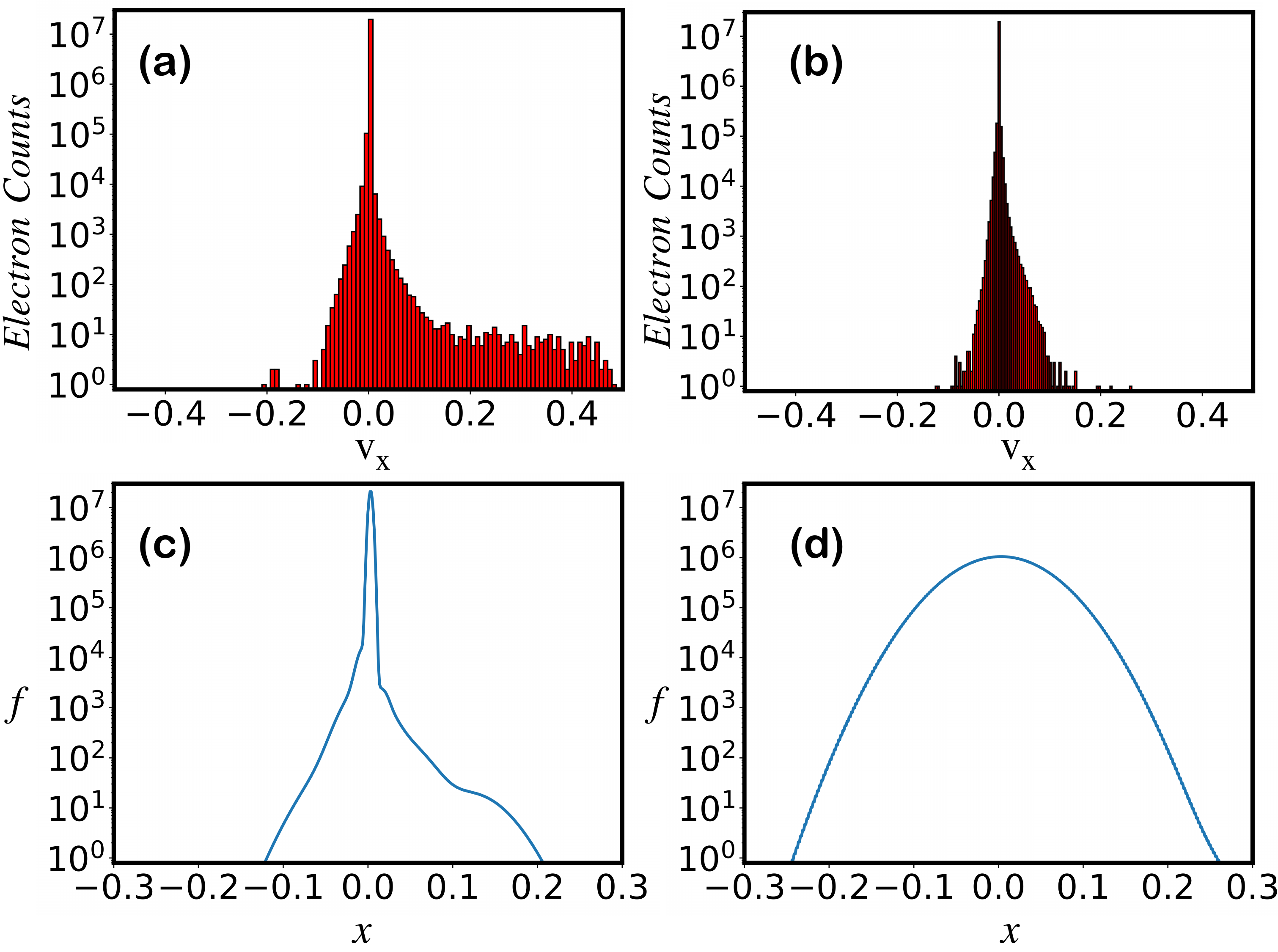}
\caption{\label{Diffusion_Vacuum} Diffusion in case of Vacuum heating. Snapshots (a) and (b) show the $\mathrm{v_x}$ component of velocity distributions at times  $t_1 = 300$  and  $t_2 = 900$, respectively, from PIC simulations. Diffusion of initial distribution profile to single maxwellian from fluid simulations have been shown in snapshots (c) and (d).}
\end{figure}

\section{Diffusion}
\label{diffusion}
In this section, we study the possible thermalization of the particle energy acquired by the laser field. Subsequently, it appears to happen through a diffusive process in the velocity space as we now illustrate. Subplots (a) and (b) in Fig. \ref{Diffusion_JxB} shows the distribution function of the $\hat{x}$ component of velocity ($\mathrm{v_x}$) at times  $t_1 = 320$  and  $t_2 = 3500$ respectively, obtained from PIC simulations for the $\vec{J}\times\vec{B}$ heating case. These results indicate that the distribution function which was initially considerably non-Maxwellian tries to approach a Maxwellian form with time. We have also modeled the initial distribution by a form shown in subplot (c)  of the figure. This form was chosen to evolve with the help of a diffusion equation in velocity space. 
The LCPFCT \cite{boris1993lcpfct} suite of subroutines was used for this purpose. It clearly relaxes to a Gaussian form shown in subplot (d). A comparison of subplots (b) and (d) shows that in the PIC simulations the higher energy particles do not seem to show up. It is likely that such particles being energetic have left the simulation box. The comparison, therefore, is not quantitatively accurate. It merely demonstrates that the velocity distribution seemingly approaches a Maxwellian form with time.  
A similar analysis was performed for the vacuum heating case. Here too the approach towards symmetric Maxwellian form can be observed in Fig. \ref{Diffusion_Vacuum}.

\section{\label{conclusion}Conclusion}
The absorption of laser energy by plasma is of enormous importance and attracts wide-ranging interests. 
Novel mechanisms are being explored to enhance energy absorption \cite{vashistha2020new}. 
 While the energy absorption through collisional and resonance absorption mechanisms \cite{kruer2019physics,decker1994nonlinear,estabrook1975two,freidberg1972resonant} are well understood, the irreversible energy transfer occurring for non-resonant and collisionless cases has often been attributed to the generation of sheath electric fields.  We have carried out Particle-In-Cell (PIC) using OSIRIS4.0 \cite{hemker2000particle, fonseca2002osiris, fonseca2008one} to show that the electron trajectories in these cases become chaotic in the combined laser and sheath fields. We show that the higher the value of the Lyapunov index, the higher the energy absorption in these cases. This has been contrasted by studying a case of Electron cyclotron resonance. It is interesting to note that though the energy absorption is very significant in the resonant case, the value of the Lyapunov index is negligible. In Fig. \ref{LLE_All}, the comparison of the average LLE and its standard deviation has been shown for the three cases that we have considered. It is thus clear that the non-resonant cases relying on the generation of sheath fields take the chaotic route for energy absorption.

\begin{acknowledgments}
The authors would like to acknowledge the OSIRIS Consortium, consisting of UCLA and IST (Lisbon, Portugal), for providing access to the OSIRIS-4.0 framework, which is the work supported by the NSF ACI-1339893. AD would like to acknowledge her J C Bose fellowship grant JCB/2017/000055 and CRG/2022/002782 grant of DST. AD  would like to thank P. H. Diamond of UCSD for valuable suggestions and discussions. AD also thanks the Isaac Newton Institute for Mathematical Sciences, Cambridge, for support and hospitality during the program Anti-Diffusion in Multiphase and Active Flows (ADIW04), where work on this paper was undertaken. It was supported by  EPSRC grant no EP/R014604/1. The authors thank the IIT Delhi HPC facility for computational resources. Rohit Juneja thanks the Council for Scientific and Industrial Research (Grant no. 09/086(1448)/2020-EMR-I) for funding the research.
\end{acknowledgments}

\section*{References}
\bibliography{aipsamp}

\end{document}